\documentclass[pre,aps,amssymb,amsmath,showpacs,superscriptaddress,twocolumn]{revtex4}

\usepackage{amsmath}
\usepackage{amssymb}
\usepackage{epsfig}
\usepackage{amscd}
\usepackage{units}
\usepackage{amsfonts}
\usepackage{graphicx,psfrag,xspace}
\usepackage{float}
\usepackage{booktabs}
\usepackage[normalem]{ulem}
\newcommand{\ind}{1\hspace{-1.3mm}{1}}
\usepackage{bm}
\usepackage{bbm}

\begin{document}

\title{Finite-size effects and percolation properties of Poisson geometries}
\author{C.~Larmier}
\affiliation{Den-Service d'\'etudes des r\'eacteurs et de math\'ematiques appliqu\'ees (SERMA), CEA, Universit\'e Paris-Saclay, F-91191, Gif-sur-Yvette, France}
\author{E.~Dumonteil}
\affiliation{IRSN, 31 Avenue de la Division Leclerc, 92260 Fontenay aux Roses, France}
\author{F.~Malvagi}
\affiliation{Den-Service d'\'etudes des r\'eacteurs et de math\'ematiques appliqu\'ees (SERMA), CEA, Universit\'e Paris-Saclay, F-91191, Gif-sur-Yvette, France}
\author{A.~Mazzolo}
\affiliation{Den-Service d'\'etudes des r\'eacteurs et de math\'ematiques appliqu\'ees (SERMA), CEA, Universit\'e Paris-Saclay, F-91191, Gif-sur-Yvette, France}
\author{A.~Zoia}
\email{andrea.zoia@cea.fr}
\affiliation{Den-Service d'\'etudes des r\'eacteurs et de math\'ematiques appliqu\'ees (SERMA), CEA, Universit\'e Paris-Saclay, F-91191, Gif-sur-Yvette, France}

\begin{abstract}
Random tessellations of the space represent a class of prototype models of heterogeneous media, which are central in several applications in physics, engineering and life sciences. In this work, we investigate the statistical properties of $d$-dimensional isotropic Poisson geometries by resorting to Monte Carlo simulation, with special emphasis on the case $d=3$. We first analyse the behaviour of the key features of these stochastic geometries as a function of the dimension $d$ and the linear size $L$ of the domain. Then, we consider the case of Poisson binary mixtures, where the polyhedra are assigned two `labels' with complementary probabilities. For this latter class of random geometries, we numerically characterize the percolation threshold, the strength of the percolating cluster and the average cluster size.
\end{abstract}


\maketitle

\section{Introduction}
\label{intro}

Heterogeneous and disordered media emerge in several applications in physics, engineering and life sciences. Examples are widespread and concern for instance light propagation through engineered optical materials~\cite{NatureOptical, PREOptical, PREQuenched} or turbid media~\cite{davis, kostinski, clouds}, tracer diffusion in biological tissues~\cite{tuchin}, neutron diffusion in pebble-bed reactors~\cite{larsen} or randomly mixed immiscible materials~\cite{renewal}, inertial confinement fusion~\cite{zimmerman, haran}, and radiation trapping in hot atomic vapours~\cite{NatureVapours}, only to name a few. Stochastic geometries provide convenient models for representing such configurations, and have been therefore widely studied~\cite{santalo, torquato, kendall, solomon, moran, ren}, especially in relation to heterogeneous materials~\cite{torquato}, stochastic or deterministic transport processes~\cite{pomraning}, image analysis~\cite{serra}, and stereology~\cite{underwood}.

A particularly relevant class of random media is provided by the so-called Poisson geometries~\cite{santalo}, which form a prototype process of isotropic stochastic tessellations: a portion of a $d$-dimensional space is partitioned by randomly generated $(d-1)$-dimensional hyper-planes drawn from an underlying Poisson process. The resulting random geometry (i.e., the collection of random polyhedra determined by the hyper-planes) satisfies the important property that an arbitrary line thrown within the geometry will be cut by the hyper-planes into exponentially distributed segments~\cite{santalo}. In some sense, the exponential correlation induced by Poisson geometries represents perhaps the simplest model of `disordered' random fields, whose single free parameter (i.e., the average correlation length) can be deduced from measured data~\cite{mikhailov}. Following the pioneering works by Goudsmit~\cite{goudsmit}, Miles~\cite{miles1964a, miles1964b} and Richards~\cite{richards} for $d=2$, the statistical features of the Poisson tessellations of the plane have been extensively analysed, and rigorous results have been proven for the limit case of domains having an infinite size: for a review, see, e.g.,~\cite{santalo, moran, ren}. An explicit construction amenable to Monte Carlo simulations for two-dimensional homogeneous and isotropic Poisson geometries of finite size has been established in~\cite{switzer}.

Theoretical results for infinite Poisson geometries have been later generalized to $d=3$, which is key for real-world applications but has comparatively received less attention, and higher dimensions by several authors~\cite{miles1969, miles1970, miles1971, miles1972, matheron, santalo}. The two-dimensional construction for isotropic Poisson geometries has been analogously extended to three-dimensional (and in principle $d$-dimensional) domains~\cite{serra, mikhailov}.

In this work, we will numerically investigate the statistical properties of $d$-dimensional isotropic Poisson geometries by resorting to Monte Carlo simulation, with special emphasis on the case $d=3$. Our aim is two-fold: first, we will focus on finite-size effects and on the convergence towards the limit behaviour of infinite domains. In order to assess the impact of dimensionality on the convergence patterns, comparisons to analogous numerical or exact findings obtained for $d=1$ and $d=2$ (where available) will be provided. In so doing, we will also present and discuss the simulation results for some physical observables for which exact asymptotic results are not known, yet.

Then, we will consider the case of `coloured' Poisson geometries, where each polyhedron is assigned a label with a given probability. Such models emerge, for instance, in connection to particle transport problems, where the label defines the physical properties of each polyhedron~\cite{pomraning, mikhailov}. The case of random binary mixtures, where only two labels are allowed, will be examined in detail. In this context, we will numerically determine the statistical features of the coloured polyhedra, which are obtained by regrouping into clusters the neighbouring volumes by their common label. Attention will be paid in particular to the percolation properties of such binary mixtures for $d=3$: the percolation threshold at which a cluster will span the entire geometry, the average cluster size and the probability that a polyhedron belongs to the the spanning cluster will be carefully examined and contrasted to the case of percolation on lattices~\cite{percolation_book}. The effect of dimensionality will be again assessed by comparison with the case $d=2$, for which analogous results were numerically determined in~\cite{lepage}.

This paper is structured as follows: in Sec.~\ref{construction} we will recall the explicit construction for $d$-dimensional isotropic Poisson geometries, with focus on $d=3$. In Sec.~\ref{uncolored_geo} we will discuss the statistical properties of Poisson geometries, and assess the convergence to the limit case of infinite domains. In Sec.~\ref{colored_geo} we will extend our analysis to the case of coloured geometries and related percolation properties. Conclusions will be finally drawn in Sec.~\ref{conclusions}. 

\section{Construction of Poisson geometries}
\label{construction}

For the sake of completeness, in this Section we will recall the strategy for the construction of Poisson geometries, spatially restricted to a $d$-dimensional box. The case $d=1$ simply stems from the Poisson point process on the line~\cite{santalo}, and will not be detailed here. The explicit construction of homogeneous and isotropic Poisson geometries for the case $d=2$ restricted to a square has been originally proposed by~\cite{switzer}, based on a Poisson point field in an auxiliary parameter space in polar coordinates. It has been recently shown that this construction can be actually extended to $d=3$ and even higher dimensions~\cite{mikhailov} by suitably generalizing the auxiliary parameter space approach of~\cite{switzer} and using the results of~\cite{serra}. In particular, such $d$-dimensional construction satisfies the homogeneity and isotropy properties~\cite{mikhailov}.

The method proposed by~\cite{mikhailov} is based on a spatial decomposition (tessellation) of the $d$-hypersphere of radius $R$ centered at the origin by generating a random number $q$ of $(d-1)$-hyperplanes with random orientation and position. Any given $d$-dimensional subspace included in the $d$-hypersphere will therefore undergo the same tessellation procedure, restricted to the region defined by the boundaries of the subspace. The number $q$ of $(d-1)$-hyperplanes is sampled from a Poisson distribution with parameter $R \Lambda_d$, with $\Lambda_d= \lambda {\cal A}_d(1)/{\cal V}_{d-1}(1) $. Here ${\cal A}_{d}(1)=2\pi^{d/2}/\Gamma(d/2)$ denotes the surface of the $d$-dimensional unit sphere ($\Gamma(a)$ being the Gamma function~\cite{special_functions}), ${\cal V}_{d}(1)=\pi^{d/2}/\Gamma(1+d/2)$ denotes the volume of the $d$-dimensional unit sphere, and $\lambda$ is the arbitrary density of the tessellation, carrying the units of an inverse length. This normalization of the density $\lambda$ corresponds to the convention used in~\cite{santalo}, and is such that $\lambda t$ yields the mean number of $(d-1)$-hyperplanes intersected by an arbitrary segment of length $t$.

\begin{figure}[t]
\begin{center}
\includegraphics[scale=0.45]{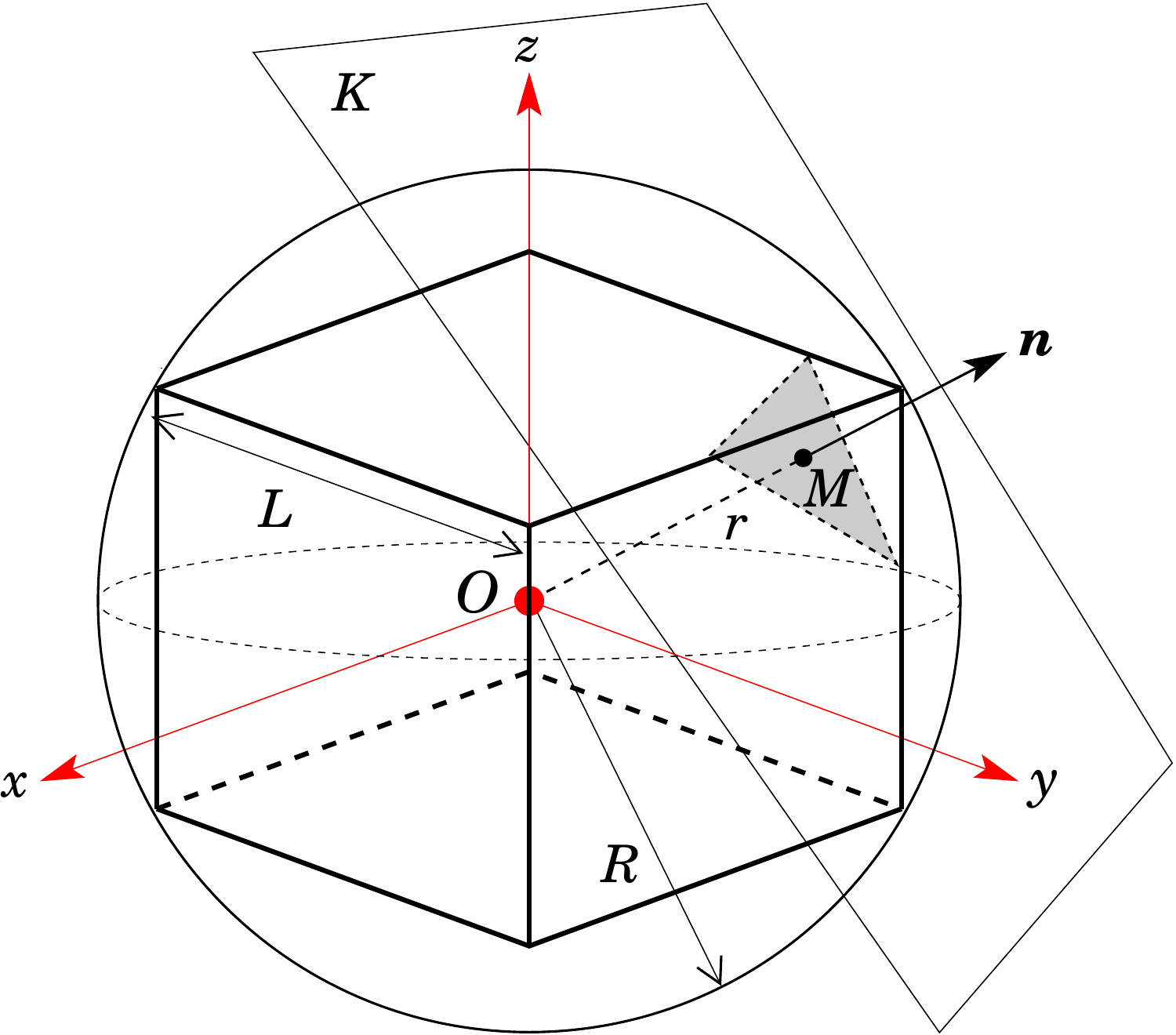}
\end{center}
\caption{(Color online) Cutting a cube with a random plane. A cube of side $L$ is centered in $O$. The circumscribed sphere centered in $O$ has a radius $R=\sqrt{3}L/2$. The point $\mathbf M$ is defined by $\mathbf M=r {\mathbf n}$, where $r$ is uniformly sampled in the interval $[0,R]$ and ${\mathbf n}$ is a random unit vector of components ${\mathbf n}=(n_1,n_2,n_3)^T$, with $n_1=1-2\xi_1$, $n_2=\sqrt{1-n_1^2}\cos{(2 \pi \xi_2)}$ and $n_3=\sqrt{1-n_1^2}\sin{(2 \pi \xi_2)}$. The auxiliary variables $\xi_1$ and $\xi_2$ are sampled from two independent uniform distributions in the interval $[0,1]$. The random plane $K$ of equation $n_1 x+n_2 y+n_3 z=r$ is orthogonal to the vector ${\mathbf n}$ and intersects the point $\mathbf M$.}
\label{fig1}
\end{figure}

Let us now focus on the case $d=3$. Suppose, for the sake of simplicity, that we want to obtain an isotropic tessellation of a box of side $L$, centered in the origin $O$. This means that the Poisson tessellation is restricted to the region $[-L/2,L/2]^3$. We denote $R$ the radius of the sphere circumscribed to the cube. The algorithm proceeds then as follows. The first step consists in sampling a random number of planes $q$ from a Poisson distribution of parameter $4 \lambda R$, where the factor $4$ stems from ${\cal A}_3(1)/{\cal V}_2(1) = 4$. The second step consists in sampling the random planes that will cut the cube. This is achieved by choosing a radius $r$ uniformly in the interval $[0,R]$ and then sampling two other random numbers, denoted $\xi_1$ and $\xi_2$, from two independent uniform distributions in the interval $[0,1]$. Based on these three random parameters, a unit vector ${\mathbf n}=(n_1,n_2,n_3)^T$ is generated (see Fig.~\ref{fig1}), with components
\begin{align}
n_1&=1-2\xi_1 \nonumber \\
n_2&=\sqrt{1-n_1^2}\cos{(2 \pi \xi_2)}\nonumber \\
n_3&=\sqrt{1-n_1^2}\sin{(2 \pi \xi_2)}.\nonumber
\end{align}
Let now $\mathbf M$ be the point such that ${\mathbf{ OM}}=r {\mathbf n}$. The random plane $K$ will be finally defined by the equation $n_1 x + n_2 y +n_3 z =r$, passing trough $\mathbf M$ and having normal vector ${\mathbf n}$. By construction, this plane does intersect the circumscribed sphere of radius $R$ but not necessarily the cube: the probability that the plane intersects both the sphere and the cube can be deduced from a classical result of integral geometry. For two convex sets $J_0$ and $J_1$ in $\mathbb R^3$, with $J_1 \subset J_0$, the probability that a randomly chosen plane meets both $J_0$ and $J_1$ is given by the ratio ${\cal M}_{1}(J_1) / {\cal M}_{1}(J_0) $, ${\cal M}_{1}(J)$ being the mean orthogonal $1$-projection of $J$ onto an isotropic random line~\cite{santalo}. The quantity ${\cal M}_{1}(J)$ takes also the name of mean caliper diameter of the set $J$~\cite{miles1972}.

The average caliper diameter of a cube of side $L$ is $3L/2$, whereas for the sphere the average caliper diameter coincides with its diameter $2R = L\sqrt{3}$, which yields a probability $\sqrt{3}/2 \simeq 0.866$ for the random planes to fall within the cube~\footnote{In the plane $\mathbb R^2$, the probability that a random line intercepts both square of side $L$ and the circumscribed circle of radius $R=\sqrt{2}L/2$ is again given by the ratio of the respective mean caliper diameters, which for $d=2$ are simply proportional to the perimeters of each set (the so-called Barbier-Crofton theorem). This yields a probability $4L/(2\pi \sqrt{2}L/2) = 2\sqrt{2}/\pi \simeq 0.900$ for a random line to fall within the square~\cite{santalo}.}.

\begin{figure}[t]
\includegraphics[scale=0.29]{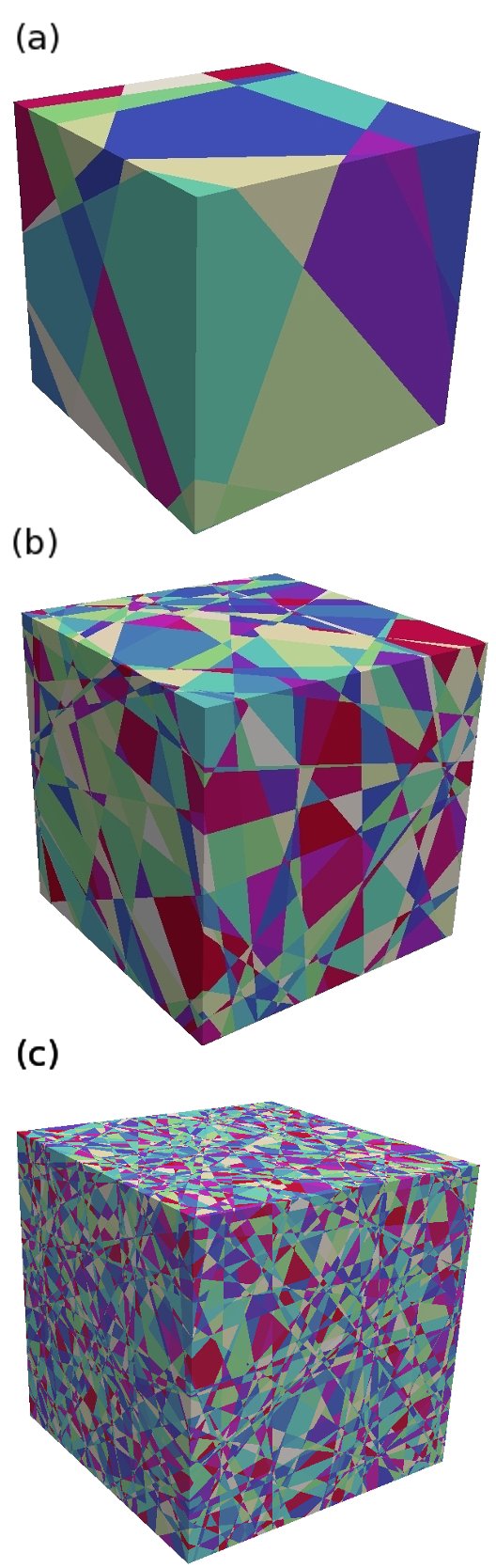}
\caption{(Color online) Examples of Monte Carlo realizations of isotropic Poisson geometries restricted to a three-dimensional box of linear size $L$. For all realizations, we have chosen $L=20$. The geometry at the top (a) has $\lambda=0.2$, that at the middle (b) has $\lambda=1$ and that at the bottom (c) has $\lambda=3$. For fixed $L$, the average number of random polyhedra composing the geometry increases with increasing $\lambda$.}
\label{fig2}
\end{figure}

The tessellation is built by successively generating the $q$ random planes. Initially, the stochastic geometry is composed of a single polyhedron, i.e., the cube. If the first sampled plane intersects the region $[-L/2,L/2]^3$, new polyhedra are generated within the cube and the tessellation is updated. This procedure is then iterated until $q$ random planes have been generated. By construction, the polyhedra defined by the intersection of such random planes are convex. For illustration purposes, some examples of isotropic Poisson tessellation of a cube of side $L=20$ obtained by Monte Carlo simulation are presented in Fig.~\ref{fig2}, for different values of the density $\lambda$. The number of random polyhedra of the tessellation increases with increasing $\lambda$.

\section{Monte Carlo analysis}
\label{uncolored_geo}

The physical observables of interest associated to the stochastic geometries, such as for instance the volume of a polyhedron, its surface, the number of edges, and so on, are clearly random variables, whose statistical distribution we would like to characterize. In the following, we will focus on the case of Poisson geometries restricted to a $d$-dimensional box of linear size $L$.

With a few remarkable exceptions, the exact distributions for the physical observables are unfortunately unknown~\cite{santalo}. A number of exact results have been nonetheless established for the (typically low-order) moments of the observables and for their correlations, at least in the limit case of domains having an infinite extension~\cite{santalo, kendall, solomon}. Monte Carlo simulation offers a unique tool for the numerical exploration of the statistical features of Poisson geometries. In particular, by resorting to the algorithm described above we can $i)$ investigate the convergence of the moments and distributions of arbitrary physical observables to their known limit behaviour (if any), and $ii)$ numerically explore the scaling of the moments and the distributions for which exact asymptotic results are not yet available. We will thus address these issues with the help of a Monte Carlo code developed to this aim.

\subsection{Number of polyhedra}

\begin{figure}[t]
\begin{center}
\includegraphics[scale=0.68]{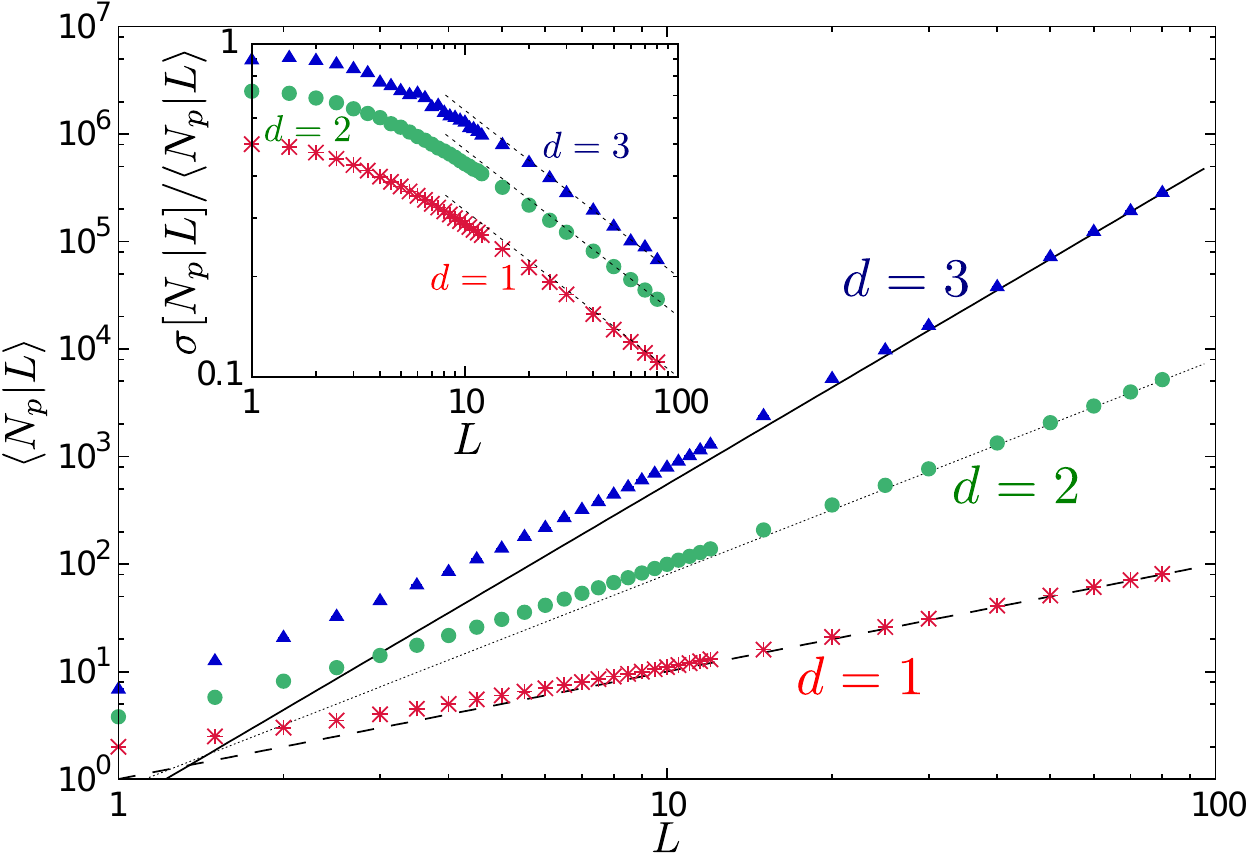}
\end{center}
\caption{(Color online) The average number $\langle N_p|L \rangle$ of $d$-polyhedra in $d$-dimensional Poisson geometries as a function of the linear size $L$ of the domain. The scaling law $ L^d$ is displayed for reference with dashed lines. Inset. The dispersion factor $\sigma[N_p|L]/\langle N_p|L \rangle$ as a function of $L$. The scaling law $1/\sqrt{L}$ is displayed for reference with dashed lines.}
\label{fig3}
\end{figure}

To begin with, we will first analyse the growth of the number $N_p$ of polyhedra in $d$-dimensional Poisson geometries as a function of the linear size $L$ of the domain, for a given value of the density $\lambda$. In the following, we will always assume that $\lambda=1$, unless otherwise specified (with both $\lambda$ and $L$ expressed in arbitrary units). The quantity $N_p$ provides a measure of the complexity of the resulting geometries. The simulation findings for the average number $\langle N_p|L\rangle$ of $d$-polyhedra (at finite $L$) and the dispersion factor, i.e., the ratio $\sigma[N_p|L]/\langle N_p|L \rangle$, $\sigma$ denoting the standard deviation, are illustrated in Fig.~\ref{fig3}. For large $L$, we find an asymptotic scaling law $\langle N_p |L\rangle \sim L^d$: the complexity of the random geometries increases with system size and dimension (Fig.~\ref{fig3}, top), as expected on physical grounds. This means that the computational cost to generate a realization of a Poisson geometry is also an increasing function of the system size and of the dimension. As for the dispersion factor, an asymptotic scaling law $\sigma[N_p|L]/\langle N_p|L\rangle \sim 1/\sqrt{L}$ is found for large $L$, independent of the dimension (Fig.~\ref{fig3}, bottom): for large systems, the distribution of $N_p$ will be then peaked around the average value $\langle N_p|L\rangle$.

\subsection{Markov properties}

Poisson geometries are Markovian, which means that in the limit case of infinite domains an arbitrary line will be cut by the $(d-1)$-surfaces of the $d$-polyhedra into segments whose lengths $\ell$ are exponentially distributed, i.e.,
\begin{equation}
{\cal P}(\ell)  = \mu e^{-\mu \ell},
\label{asy_length}
\end{equation}
with average density $\mu = \lambda$. Conversely, the number of intersections $n_i$ of an arbitrary segment of length $t$ with the $(d-1)$-surfaces of the $d$-polyhedra in an infinite domain will obey a Poisson distribution
\begin{equation}
{\cal P}(n_i) = \nu^{n_i} \frac{e^{-\nu}}{n_i!},
\label{asy_poisson}
\end{equation}
with mean value $\nu = \lambda t$.

In order to verify that the geometries constructed by resorting to the algorithm described above satisfy the Markov properties, we have numerically computed by Monte Carlo simulation the probability density of the segment lengths and the probability of the number of intersections as a function of the linear size $L$ of the domain and for different dimensions $d$. For the former, a Poisson geometry is first generated, and a line is then drawn by uniformly choosing a point in the box and an isotropic direction: this choice corresponds to formally assuming a so-called $I$-randomness for the lines~\cite{coleman}. The intersections of the line with the polyhedra of the geometry are computed, and the resulting segment lengths are recorded. The whole procedure is repeated a large number of times in order to get the appropriate statistics. For the latter, a test segment of unit length is sampled by choosing a point and a direction as before, and the number of intersections with the polyhedra are again determined.

\begin{figure}[t]
\begin{center}
\includegraphics[scale=0.68]{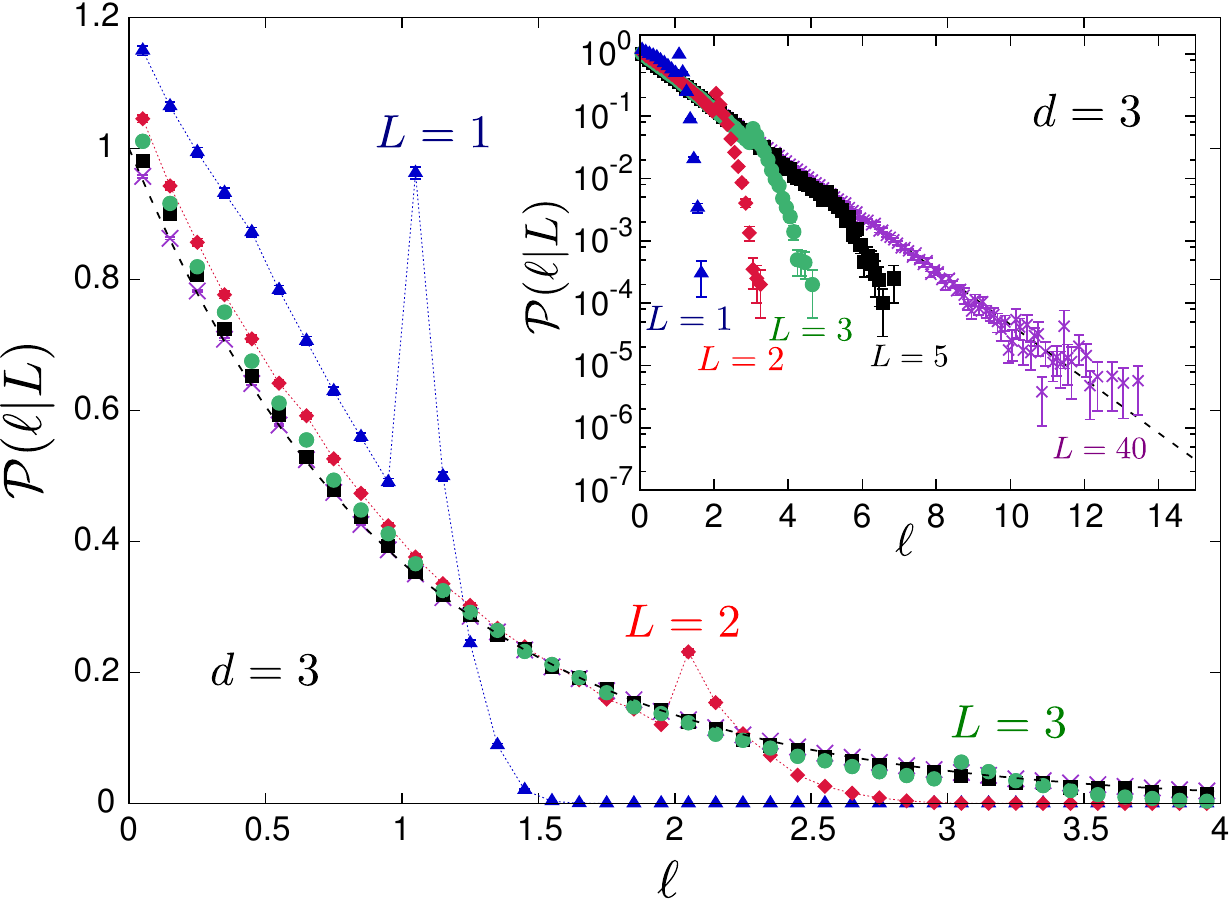}
\end{center}
\caption{(Color online) The probability densities ${\cal P}(\ell|L)$ of the segment lengths as a function of the linear size $L$ of the domain, in dimension $d=3$. Symbols correspond to the Monte Carlo simulation results, with lines added to guide the eye: blue triangles denote $L=1$, red diamonds $L=2$, green circles $L=3$, black squares $L=5$, and purple crosses $L=40$. The asymptotic (i.e., $L \to \infty$) exponential distribution given in Eq.~\eqref{asy_length} is displayed as a black dashed line for reference. The inset displays the same data in log-linear scale.}
\label{fig4}
\end{figure}

\begin{figure}[t]
\begin{center}
\includegraphics[scale=0.68]{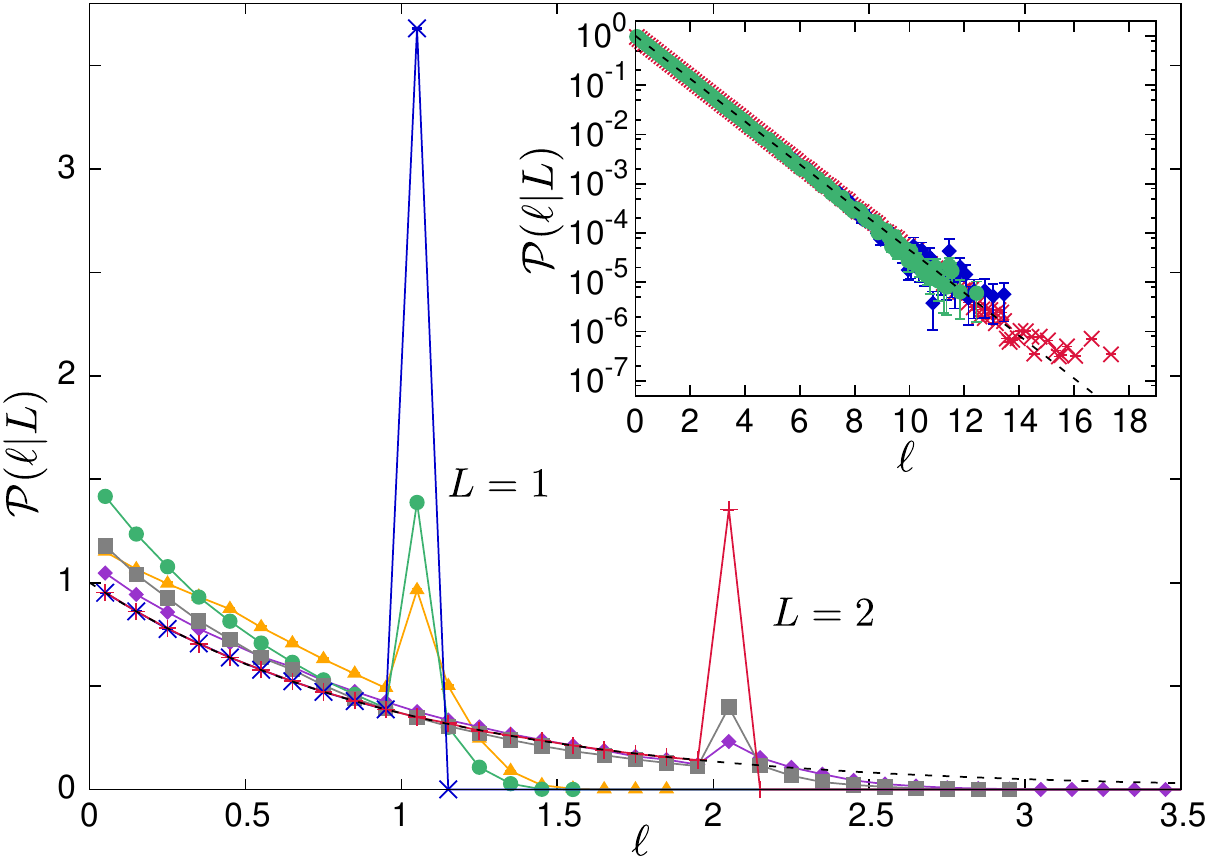}
\end{center}
\caption{(Color online) The probability densities ${\cal P}(\ell|L)$ of the segment lengths as a function of the linear size $L$ of the domain and of the dimension $d$. Symbols correspond to the Monte Carlo simulation results, with lines added to guide the eye: for $L=1$, blue crosses denote $d=1$, green circles $d=2$, and orange triangles $d=3$; for $L=2$, red pluses denote $d=1$, grey squares $d=2$, and purple diamonds $d=3$. The asymptotic (i.e., $L \to \infty$) exponential distribution given in Eq.~\eqref{asy_length} is displayed as a black dashed line for reference. Inset. The case of a system size $L=40$: red crosses denote $d=1$, green circles $d=2$, and blue diamonds $d=3$; the black dashed line corresponds to Eq.~\eqref{asy_length}.}
\label{fig5}
\end{figure}

The numerical results for ${\cal P}(\ell|L)$ at finite $L$ are illustrated in Figs.~\ref{fig4} and~\ref{fig5}. For small $L$, finite-size effects are apparent in the segment length density: this is due to the fact that the longest line that can be drawn across a box of linear size $L$ is $\sqrt{d} L$, which thus induces a cut-off on the distribution (see Fig.~\ref{fig4}). For $\lambda L \gg 1$, the finite-size effects due to the cut-off fade away and the probability densities eventually converge to the expected exponential behaviour. The rate of convergence appears to be weakly dependent on the dimension $d$ (see Fig.~\ref{fig5}). The case $d=1$ can be treated analytically and might thus provide a rough idea of the approach to the limit case. For any finite $L$, the distribution of the segment lengths for $d=1$ is
\begin{equation}
{\cal P}(\ell|L) = \lambda e^{-\lambda \ell} \ind_{\ell < L} + e^{-\lambda L} \delta(\ell-L),
\end{equation}
$\ind_J$ being the marker function of the domain $J$. The moments of order $m$ of the segment length $\ell$ for finite $L$ thus yield
\begin{equation}
\langle \ell^m |L \rangle = \frac{\Gamma(m+1)}{\lambda^m}-\frac{\Gamma_{m+1}(\lambda L)}{\lambda^m} + e^{-\lambda L} L^m,
\end{equation}
where $\Gamma_a(x)$ is the incomplete Gamma function~\cite{special_functions}. In the limit case $L \to \infty$, we have $\langle \ell^m \rangle = \Gamma(m+1)/\lambda^m$, so that for the convergence rate we obtain
\begin{equation}
\frac{\langle \ell^m |L\rangle}{\langle \ell^m \rangle} = 1-\frac{\Gamma_{m+1}(\lambda L) -e^{-\lambda L} (\lambda L)^m}{\Gamma(m+1)},
\label{exact_moment_d1}
\end{equation}
which for large $\lambda L \gg 1$ gives
\begin{equation}
\frac{\langle \ell^m |L\rangle}{\langle \ell^m \rangle} \simeq 1-\frac{(\lambda L)^{m-1} e^{-\lambda L}}{\Gamma(m+1)}.
\end{equation}
Thus, the average segment length ($m=1$) converges exponentially fast to the limit behaviour, whereas the higher moments ($m \ge 2$) converge sub-exponentially with power-law corrections. For $d>1$, the cut-off is less abrupt, but the distributions ${\cal P}(\ell|L)$ still show a peak at $\ell=L$, and vanish for $\ell > L \sqrt{d}$. The asymptotic average segment lengths for $L \to \infty$ yield $\langle \ell \rangle = 1/\lambda$ for any $d$: the Monte Carlo simulation results obtained for a large $L=80$ are compared to the theoretical formulas in Tab.~\ref{tab1}.

\begin{table}[h!]
\footnotesize
\begin{ruledtabular}
\begin{tabular}{c c c c}
$d$ & $\langle \ell \rangle$ & Theoretical value & Monte Carlo \\
\hline
$1$ & $1/\lambda$ & $1$ & $1.0002 \pm 10^{-4}$ \\
$2$ & $1/\lambda$ & $1$ & $0.9932 \pm 6\times 10^{-4}$ \\
$3$ & $1/\lambda$ & $1$ & $0.9985 \pm 3 \times 10^{-3}$ \\
\end{tabular}
\end{ruledtabular}
\caption{The average segment lengths $\langle \ell \rangle$. Monte Carlo simulation results are obtained with $L=80$ and $\lambda=1$ for any dimension $d$.}
\label{tab1}
\end{table}

For $d=1$ we performed $10^6$ realizations, with an average number $\langle N_p \rangle = 80.986 \pm 9 \times 10^{-3}$ of $1$-polyhedra per realization. For $d=2$ we performed $10^5$ realizations, with an average number $\langle N_p \rangle = 5189 \pm 3$ of $2$-polyhedra per realization. For $d=3$ we performed $2 \times 10^3$ realizations, with an average number $\langle N_p \rangle = 2.82\times 10^5 \pm 1.4 \times 10^3$ $3$-polyhedra per realization.

The convergence of the distribution of the number of intersections to the limit Poisson distribution ${\cal P}(n_i)$ is very fast as a function of $L$, which most probably stems from the unit test segment being only weakly affected by finite-size effects (i.e., by the polyhedra that are cut by the boundaries of the box), contrary to the case of the lines. Finite-size effects are appreciable only for large values of the number of intersections $n_i$, which in turn occur with small probability. The asymptotic average number of intersections per unit length for $L \to \infty$ yield $\langle n_i \rangle = \lambda$ for any $d$: the Monte Carlo simulation results obtained for a large $L=80$ are compared to the theoretical formulas in Tab.~\ref{tab2}, with the same simulation parameters as above.

\begin{table}[h!]
\footnotesize
\begin{ruledtabular}
\begin{tabular}{c c c c}
$d$ & $\langle n_i \rangle$ & Theoretical value & Monte Carlo \\
\hline
$1$ & $\lambda$ & $1$ & $1.001 \pm 10^{-3}$ \\
$2$ & $\lambda$ & $1$ & $0.995 \pm 3\times 10^{-3}$ \\
$3$ & $\lambda$ & $1$ & $1.03 \pm 2 \times 10^{-2}$ \\
\end{tabular}
\end{ruledtabular}
\caption{The average number of intersections $\langle n_i \rangle$. Monte Carlo simulation results are obtained with $L=80$ and $\lambda=1$ for any dimension $d$.}
\label{tab2}
\end{table}

\subsection{The inradius distribution}

The inradius $r_\text{in}$ is defined as the radius of the largest sphere that can be contained in a (convex) polyhedron, and as such represents a measure of the linear size of the polyhedron~\cite{santalo}. The probability density of the inradius is exactly known in any dimension $d$ for Poisson geometries of infinite size: it turns out that $r_\text{in}$ has an exponential distribution, namely,
\begin{equation}
{\cal P}(r_\text{in}) = \Lambda_d e^{-\Lambda_d  r_\text{in}},
\label{asy_inradius}
\end{equation}
where the dimension-dependent constant $\Lambda_d$ reads $\Lambda_1=2 \lambda$, $\Lambda_2=\pi \lambda$, and $\Lambda_3=4 \lambda$. In principle, it would be possible to analytically determine the coordinates of the center and the radius of the largest contained sphere, once the equations of the $(d-1)$-hyperplanes defining the $d$-polyhedron are known~\cite{sahu}. We have however chosen to numerically compute the inradius by resorting to a linear programming algorithm. For a given realization of a Poisson geometry, we select in turn a convex $d$-polyhedron: this will be formally defined by a set ${\bf x} \in \mathbb{R}^d$ such that
\begin{equation}
{\bf a}_i^T {\bf x} \leq {\bf b}_i \,\,\, (1 \leq i \leq q),
\end{equation}
where $q$ is the number of $(d-1)$-hyperplanes composing the surface of the $d$-polyhedron. The inradius $r_\text{in}$ can be then computed based on the Chebyshev center $({\bf x},r_\text{in})$ of the $d$-polyhedron, which can be found by maximising $r_\text{in}$ with the constraints
\begin{align}
&\forall i \in \{1, 2, ..., q\}, \,\,\, {\bf a}_i^T {\bf x} + r_\text{in} || {\bf a}_i || \leq {\bf b}_i \\
& r_\text{in} > 0.
\end{align}
This maximisation problem has been finally solved by using the simplex method~\cite{recipes}.

\begin{figure}[t]
\begin{center}
\includegraphics[scale=0.68]{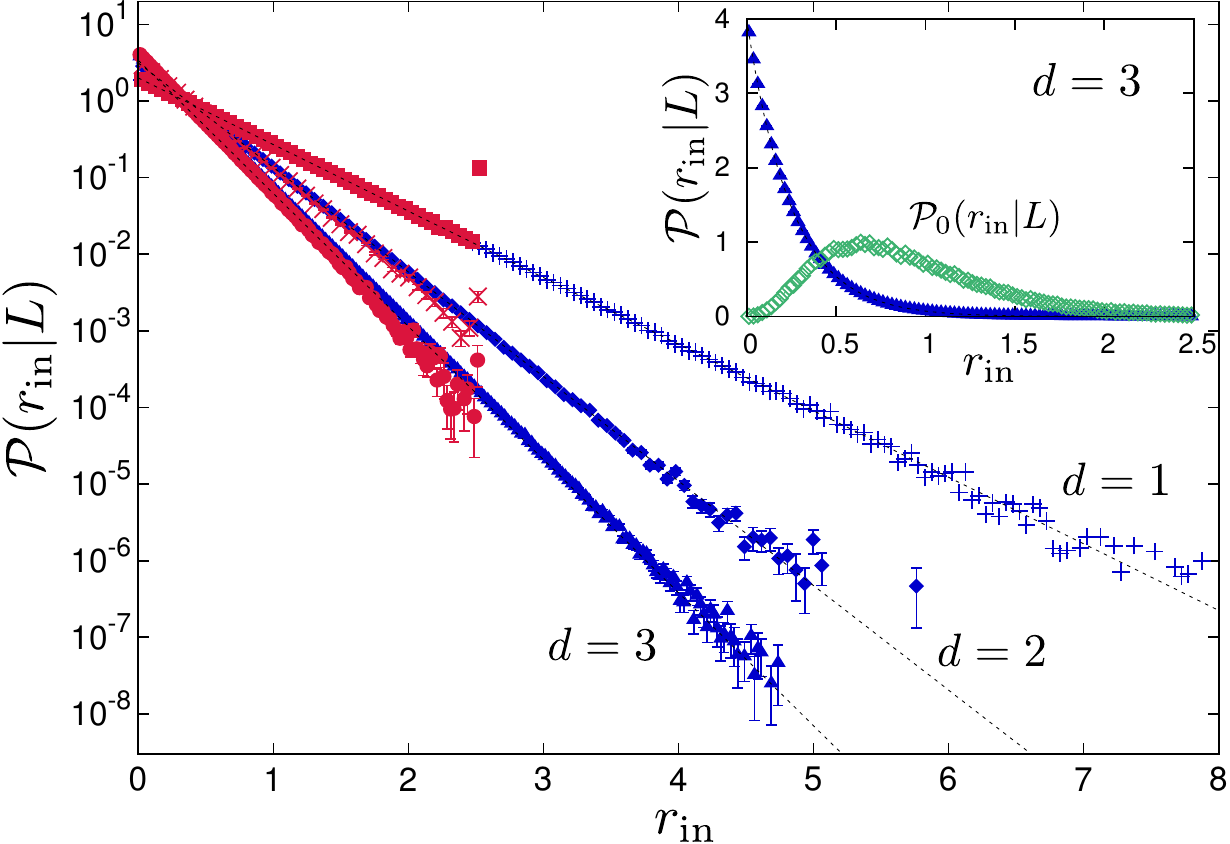}
\end{center}
\caption{(Color online) The probability density ${\cal P}(r_{in}|L)$ of the inradius as a function of the system size $L$ and of the dimension $d$. Symbols correspond to Monte Carlo simulation results. For $d=1$, red squares denote $L=5$ and blue pluses $L=40$. For $d=2$, red crosses denote $L=5$ and blue diamonds $L=40$. For $d=3$, red circles denote $L=5$ and blue triangles $L=40$. The black dashed lines represent the asymptotic (i.e., $L \to \infty$) distribution in Eq.~\eqref{asy_inradius}. Inset. Comparison between ${\cal P}(r_{in}|L)$ for a typical polyhedron (blue triangles) and ${\cal P}_0(r_{in}|L)$ for the polyhedron containing the origin (green circles), for $d=3$ and $L=40$. The dashed line represents the asymptotic distribution in Eq.~\eqref{asy_inradius}.}
\label{fig6}
\end{figure}

The results of the Monte Carlo simulation for $r_\text{in}$ are shown in Fig.~\ref{fig6} as a function of $L$ and $d$. The case $d=1$ is straightforward, since the inradius simply coincides with the half-length of the $1$-polyhedron. For any finite $L$, the numerical distributions suffer from finite-size effects, analogous to those affecting the distributions of the segment lengths $\ell$: in particular, a cut-off appears at $r_\text{in}=L/2$. As $\lambda L \gg 1$, finite-size effects fade away and the numerical distributions converge to the expected exponential behaviour. The convergence rate as a function of the system size $L$ is weakly dependent on the dimension $d$. The asymptotic average inradius for $L \to \infty$ yields $\langle r_\text{in} \rangle = 1/\Lambda_d $: the Monte Carlo simulation results obtained for a large $L=80$ are compared to the theoretical formulas in Tab.~\ref{tab3}, with the same simulation parameters as above.

\begin{table}[b]
\footnotesize
\begin{ruledtabular}
\begin{tabular}{c c c c}
$d$ & $\langle r_{\text{in}} \rangle$ & Theoretical value & Monte Carlo \\
\hline
$1$ & $1/2 \lambda$ & $0.5$ & $0.50009 \pm 6 \times 10^{-5}$ \\
$2$ & $1/\pi \lambda$ & $0.31831$ & $0.31795 \pm 9 \times 10^{-5}$ \\
$3$ & $1/4 \lambda$ & $0.25$ & $0.2499 \pm 4 \times 10^{-4}$ \\
\end{tabular}
\end{ruledtabular}
\caption{The average inradius $\langle r_{\text{in}} \rangle$. Monte Carlo simulation results are obtained with $L=80$ and $\lambda=1$ for any dimension $d$.}
\label{tab3}
\end{table}

\begin{figure}[t]
\begin{center}
\includegraphics[scale=0.68]{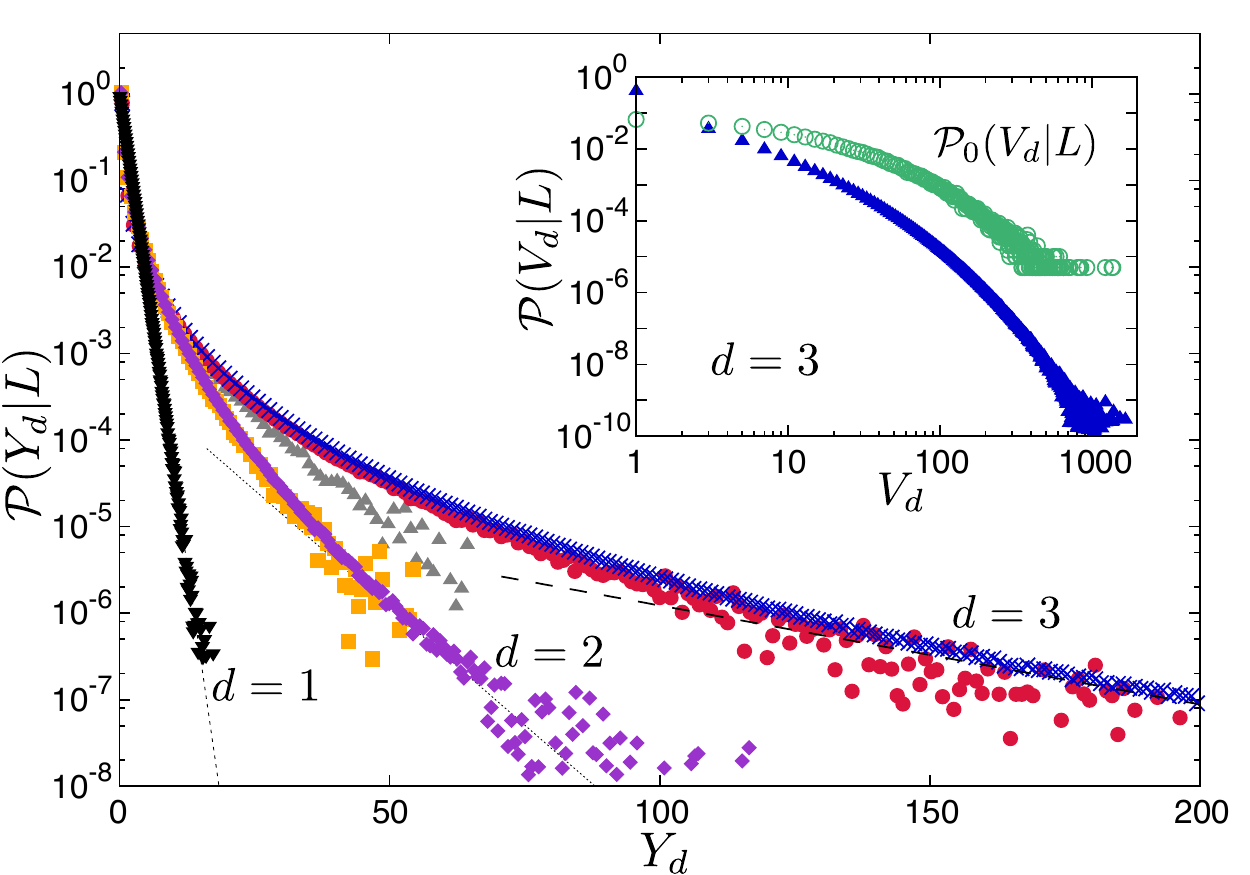}
\end{center}
\caption{(Color online) The probability density ${\cal P}(Y_d|L)$ of the dimensionless $d$-volume $Y_d=V_d / \langle V_d \rangle$ as a function of the linear size $L$ of the domain and of the dimension $d$. Black inverted triangles denote a system size $L=40$ for $d=1$. For $d=2$, purple diamonds are chosen for a system size $L=40$ and orange squares for $L=10$. For $d=3$, blue crosses are chosen for a system size $L=40$, red circles for $L=10$ and grey triangles for $L=5$. For $d=1$, the asymptotic (i.e., $L \to \infty$) exponential distribution given in Eq.~\eqref{asy_length} is displayed as a black dashed line. For $d=2$ and $d=3$, dashed lines denote exponential decay. Inset. Comparison between ${\cal P}(V_d|L)$ for a typical polyhedron (blue triangles) and ${\cal P}_0(V_d|L)$ for the polyhedron containing the origin (green circles), for $d=3$.}
\label{fig7}
\end{figure}

\subsection{The volume distribution}

One of the most important physical observables related to the stochastic geometries is the distribution ${\cal P}(V_d)$ of the $d$-volumes $V_d$ of the polyhedra. For $d=1$, this distribution coincides with that of the segment lengths, ${\cal P}(\ell)$, which means that the approach to the limit case of infinite domains follows from the same arguments as above. Unfortunately, the functional form of the distribution ${\cal P}(V_d)$ is not known for $d>1$~\cite{miles1971, miles1972, santalo}. We have thus resorted to Monte Carlo simulation so as to assess the impact of the domain size $L$ and of the dimension $d$ on ${\cal P}(V_d | L)$ for finite $L$. In order to compare the results for different $d$, we found convenient to introduce the dimensionless variable $Y_d = V_d/\langle V_d\rangle$, where the asymptotic average $d$-volume size is estimated by Monte Carlo for large $L$. The numerical findings are shown in Fig.~\ref{fig7}. It is apparent that for $\lambda L \gg 1$ the distributions ${\cal P}(Y_d | L)$ approach an asymptotic shape. The rate of convergence as a function of $L$ decreases with increasing $d$, which is expected on physical grounds because the complexity of the geometries grows as $\sim L^d$. The tails of ${\cal P}(Y_d)$ for large values of the argument $Y_d$ also depend on $d$: for $d=1$, ${\cal P}(Y_d) \sim \exp(-Y_d)$, whereas for $d>1$ the tail appears to be increasingly slower as a function of $d$. Due to poor statistics for very large values of $Y_d$, we are not able to precisely characterize the asymptotic decay of ${\cal P}(Y_d)$. It seems however that for $d>1$ the tail is not purely exponential, and that power law corrections might thus appear.

\begin{figure}[t]
\begin{center}
\includegraphics[scale=0.68]{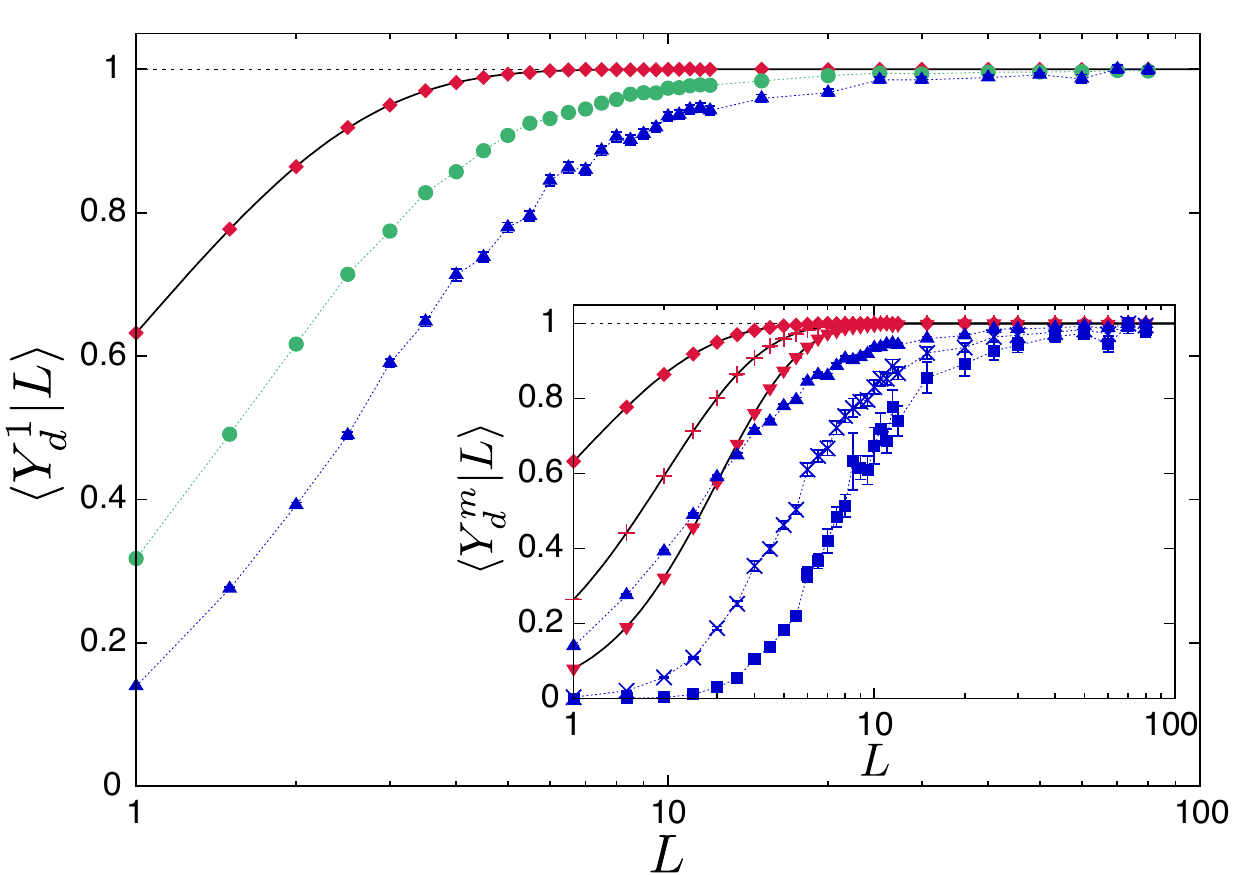}
\end{center}
\caption{(Color online) The dimensionless first moment $\langle Y^1_d |L\rangle=\langle V^1_d |L\rangle /\langle V^1_d\rangle$ of the $d$-volume, as a function of the system size $L$ and of the dimension $d$. Monte Carlo simulation results are displayed as symbols, with dashed lines lines to guide the eye for $d=2$ and $d=3$. For $d=1$, the solid line represents the exact formula given in Eq.~\eqref{exact_moment_d1}. Red diamonds denote $d=1$; green circles denote $d=2$; blue triangles denote $d=3$. Inset. The dimensionless moments $\langle Y^m_d |L\rangle=\langle V^m_d |L\rangle /\langle V^m_d\rangle$ of the $d$-volume, for $m=1,2,3$, as a function of the system size $L$, for $d=1$ and $d=2$. Monte Carlo simulation results are displayed as symbols, with dashed lines lines to guide the eye for $d=3$. For $d=1$, the solid line represents the exact formula given in Eq.~\eqref{exact_moment_d1}. For $d=1$, red diamonds denote $m=1$; red pluses denote $m=2$; red inverted triangles denote $m=3$. For $d=3$, blue triangles denote $m=1$; blue crosses denote $m=2$; blue squares denote $m=3$.}
\label{fig8}
\end{figure}

Supplementary information can be retrieved from the analysis of the $m$-th moments $\langle V_d^m\rangle$, for which exact results are available in the case $m=1,2$ and $3$ for infinite domains~\cite{miles1971, miles1972, santalo, matheron}. The convergence of the dimensionless moments $\langle Y^m_d|L\rangle=\langle V^m_d |L\rangle /\langle V^m_d\rangle$ to the limit case as a function of $L$ is displayed Fig.~\ref{fig8}. The convergence to the asymptotic value $\lim_{L\to \infty} \langle Y^m_d|L\rangle =1$ is increasingly slower as a function of $L$ as $d$ increases, whereas the order $m$ of the moments has a weak impact on the convergence rate. The Monte Carlo simulation results for the asymptotic $m$-th moments $\langle V_d^m\rangle$ obtained for a large $L=80$ are finally compared to the theoretical formulas in Tab.~\ref{tab4} for $\langle V_d \rangle$, in Tab.~\ref{tab5} for $\langle V^2_d \rangle$, and in Tab.~\ref{tab6} for $\langle V^3_d \rangle$, respectively, with the same simulation parameters as above.

\begin{table}[h!]
\footnotesize
\begin{ruledtabular}
\begin{tabular}{c c c c}
$d$ & $\langle V_d \rangle$ & Theoretical value & Monte Carlo \\
\hline
$1$ & $1/\lambda$ & $1$ & $1.0002 \pm 10^{-4}$ \\
$2$ & $4/\pi \lambda^2$ & $1.27324$ & $1.2703 \pm 7 \times 10^{-4}$ \\
$3$ & $6/\pi \lambda^3$ & $1.90986$ & $1.91 \pm 10^{-2}$ \\
\end{tabular}
\end{ruledtabular}
\caption{The average $d$-volume size $\langle V_d \rangle$. Monte Carlo simulation results are obtained with $L=80$ and $\lambda=1$ for any dimension $d$.}
\label{tab4}
\end{table}

\begin{table}[h!]
\begin{ruledtabular}
\footnotesize
\begin{tabular}{c c c c}
$d$ & $\langle V_d^2 \rangle$ & Theoretical value & Monte Carlo \\
\hline
$1$ & $2/\lambda^2$ & $2$ & $2.0007 \pm 5\times 10^{-4}$ \\
$2$ & $8/\lambda^4$ & $8$ & $7.9609 \pm 9 \times 10^{-4}$ \\
$3$ & $48/\lambda^6$ & $48$ & $47.7 \pm 0.5$ \\
\end{tabular}
\end{ruledtabular}
\caption{The second moment $\langle V^2_d \rangle$ of the $d$-volume. Monte Carlo simulation results are obtained with $L=80$ and $\lambda=1$ for any dimension $d$.}
\label{tab5}
\end{table}

\begin{table}[h!]
\footnotesize
\begin{ruledtabular}
\begin{tabular}{c c c c}
$d$ & $\langle V_d^3 \rangle$ & Theoretical value & Monte Carlo \\
\hline
$1$ & $6/\lambda^3$ & $6$ & $6.003 \pm 3 \times 10^{-3}$ \\
$2$ & $256 \pi/7\lambda^6$ & $114.893$ & $114.1 \pm 0.2$ \\
$3$ & $1344 \pi/\lambda^9$ & $4222.3$ & $4144 \pm 75$ \\
\end{tabular}
\end{ruledtabular}
\caption{The third moment $\langle V^3_d \rangle$ of the $d$-volume. Monte Carlo simulation results are obtained with $L=80$ and $\lambda=1$ for any dimension $d$.}
\label{tab6}
\end{table}

\subsection{The moments of the surfaces}

The analysis of the $d$-surfaces $A_d$ of the $d$-polyhedra is also of utmost importance, in that it provides information on the interface between the constituents of the geometry (see for instance the considerations in~\cite{miles1972}). We have then computed the first few moments $\langle A^m_d \rangle$ of the $d$-surfaces by Monte Carlo simulation. Results are recalled in Tab.~\ref{tab7}, where we compare the numerical findings for large $L=80$ to the exact formulas for infinite domains.

\begin{table}[h!]
\footnotesize
\begin{ruledtabular}
\begin{tabular}{c c c c}
 & $\langle A^m_d \rangle$ & Theoretical value & Monte Carlo \\
\hline
$\langle A_2 \rangle$ & $4/\lambda$ & $4$ & $3.995 \pm 10^{-3}$ \\
$\langle A_2^2 \rangle$ & $(2\pi^2+8)/\lambda^2$ & $27.74$ & $27.67 \pm 2 \times 10^{-2}$ \\
\colrule
$\langle A_3 \rangle$ & $24/\pi \lambda^2$ & $7.64$ & $7.63 \pm 2 \times 10^{-2} $ \\
$\langle A_3^2 \rangle$ & $240/\lambda^4$ & $240$ & $239.5 \pm 1.7$ \\
\end{tabular}
\end{ruledtabular}
\caption{The moments $\langle A^m_d\rangle$ of the $d$-surface of the $d$-polyhedra. Monte Carlo simulation results are obtained with $L=80$ and $\lambda=1$ for any dimension $d$.}
\label{tab7}
\end{table}

\subsection{The moments of the outradius}

The outradius $r_\text{out}$ is defined as the radius of the smallest sphere enclosing a (convex) polyhedron, and can be thus used together with the inradius so as to characterize the shape of the polyhedra. For $d=1$, the outradius coincides with the inradius. The probability density and the moments of the outradius of Poisson geometries for $d>1$ are not known. We have then numerically computed the moments of the outradius by resorting to an algorithm recently proposed in~\cite{fischer}. This algorithm implements a pivoting scheme similar to the simplex method for linear programming. It starts with a large $d$-ball that includes all vertices of the convex $d$-polyhedron and progressively shrinks it~\cite{fischer}. For reference, the Monte Carlo simulation results for the first few moments of $r_\text{out}$ obtained for a large $L=80$ are given in Tab.~\ref{tab8}, with the same simulation parameters as above: these numerical findings might inspire future theoretical advances.

\begin{table}[h!]
\footnotesize
\begin{tabular}{ c @{\qquad} c @{\qquad} c }
\toprule
\botrule
 $d$ & & Monte Carlo \\
\colrule
$2$ & $\langle r_\text{out} \rangle$ & $0.8444	\pm 2 \times 10^{-4}$ \\
$2$ & $\langle r_\text{out}^2 \rangle$ & $1.2291	\pm 7 \times 10^{-4}$ \\
\colrule
$3$ & $\langle r_\text{out} \rangle$ & $1.153 \pm 2 \times 10^{-3}$ \\
$3$ & $\langle r_\text{out}^2 \rangle$ & $2.127 \pm 7 \times 10^{-3}$ \\
\botrule
\end{tabular}
\caption{The moments $\langle r^m_\text{out} \rangle$ of the outradius in dimension $d$. Monte Carlo simulation results are obtained with $L=80$ and $\lambda=1$ for any dimension $d$.}
\label{tab8}
\end{table}

\subsection{The polyhedron containing the origin}

So far, the properties of the constituents of the Poisson geometries have been derived by assuming that each $d$-polyhedron has an identical statistical weight (for a precise definition, see, e.g.,~\cite{miles1964a, miles1970, miles1971, matheron}). It is also possible to attribute to each $d$-polyhedron a statistical weight equal to its $d$-volume. It can be shown that the statistics of any observable related to the $d$-polyhedron containing the origin $O$ obeys this latter volume-weighted distribution~\cite{miles1970}. This surprising property can be understood by following the heuristic argument proposed by Miles~\cite{miles1964a}: the origin has greater chances of falling within a larger rather than a smaller volume. In particular, for the moments $\langle X \rangle_0$ of the $d$-polyhedron containing the origin we formally have
\begin{equation}
\langle X \rangle_0 = \frac{\langle V_d X \rangle}{\langle V_d \rangle},
\end{equation}
where $X$ denotes an arbitrary observable~\cite{miles1970}. We have carried out an extensive analysis of the moments of the features of the $d$-polyhedra containing the origin by Monte Carlo simulation: numerical findings for the most relevant quantities are reported in Tab.~\ref{tab9}. For some of the computed quantities, such as the average inradius $\langle r_\text{in} \rangle_0$ or the average outradius $\langle r_\text{out} \rangle_0$, exact results are not available, and our numerical findings may thus support future theoretical investigations.

The full distribution ${\cal P}_0(r_\text{in}|L)$ of the inradius of the $d$-polyhedron containing the origin has been estimated, and is compared to ${\cal P}(r_\text{in}|L)$ for the inradius of a typical polyhedron of the tessellation in the inset of Fig.~\ref{fig6} for $d=3$ and a large system size $L=40$: it is immediately apparent that $\langle r_\text{in} \rangle_0 > \langle r_\text{in} \rangle$. Moreover, the behaviour of the two distributions for small $r_\text{in}$ is also different: for $L \to \infty$, ${\cal P}(r_\text{in}|L)$ attains a finite value for $r_\text{in} \to 0$ due to its exponential shape; on the contrary, our Monte Carlo simulations seem to suggest a power-law scaling ${\cal P}_0(r_\text{in}|L) \sim r^{\alpha_d}_\text{in}$ for $r_\text{in} \to 0$, with $\alpha_d = 1+(d-1)/2$.

The distribution ${\cal P}_0(V_d|L)$ of the $d$-volume of the $d$-polyhedron containing the origin has been also computed, and is compared to ${\cal P}(V_d|L)$ for the $d$-volume of a typical polyhedron of the tessellation in the inset of Fig.~\ref{fig7} for $d=3$ and a large system size $L=40$. Again, $\langle V_d \rangle_0 > \langle V_d \rangle$.

\begin{table}[h!]
\footnotesize
\begin{ruledtabular}
\begin{tabular}{c c c c c}
$d$ & & Formula & Theoretical value & Monte Carlo \\
\hline
$1$ & $\langle V_1 \rangle_0$ & $2/ \lambda$ & $2$ & $2.000 \pm 10^{-3} $ \\
$1$ & $\langle V_1^2 \rangle_0$ & $6/\lambda^2$ & $6$ & $6.001 \pm 9 \times 10^{-3}
$ \\
\hline
$2$ & $\langle V_2 \rangle_0$ & $2 \pi / \lambda^2$ & $6.28319$ & $6.28 \pm 2 \times 10^{-2}$ \\
$2$ & $\langle V_2^2 \rangle_0$ & $64 \pi^2 /7 \lambda^4$ & $90.2364$ & $90.6	\pm 0.9$ \\
$2$ & $\langle A_2 \rangle_0$ & $\pi^2/ \lambda$ & $9.8696$ & $9.87	\pm 2 \times 10^{-2}$ \\
$2$ & $\langle r_\text{in} \rangle_0$ & & & $0.886 \pm 2 \times 10^{-3}$ \\
$2$ & $\langle r_\text{out} \rangle_0$ & & & $2.028 \pm 3 \times 10^{-2}$ \\
\hline
$3$ & $\langle V_3 \rangle_0$ & $8 \pi / \lambda^3$ & $25.1327$ & $25.3 \pm 0.9$ \\
$3$ & $\langle V_3^2 \rangle_0$ & $224 \pi^2 / \lambda^6$ & $2210.79$ & $2129.1 \pm 182$ \\
$3$ & $\langle A_3 \rangle_0$ & $16 \pi / \lambda^2$ & $50.2655$ & $50.6 \pm 1.0$ \\
$3$ & $\langle r_\text{in} \rangle_0$ & & & $0.89 \pm 10^{-2}$ \\
$3$ & $\langle r_\text{out} \rangle_0$ & & & $3.11 \pm 3 \times 10^{-2}$ \\
\end{tabular}
\end{ruledtabular}
\caption{Moments of the $d$-polyhedron containing the origin. Monte Carlo simulation results are obtained with $L=80$ and $\lambda=1$ for any dimension $d$.}
\label{tab9}
\end{table}

\subsection{Other moments and correlations}

A number of moments and correlations of other physical observables are exactly known for Poisson geometries of infinite size for $d=2$ and $d=3$. For the sake of completeness, our Monte Carlo estimates corresponding to these quantities are reported in Appendix~\ref{appendix_moments}. When analytical results are not known, Monte Carlo simulation findings are displayed for reference.

\section{Coloured geometries}
\label{colored_geo}

So far, we have addressed the statistical properties of Poisson geometries based on the assumption that all polyhedra share the same physical properties, i.e., the medium is homogeneous. In many applications, the polyhedra emerging from a random tessellation are actually characterized by different physical properties, which for the sake of simplicity can be assumed to be piece-wise constant over each volume. Such stochastic mixtures can be then formally described by assigning a distinct `label' (also called `color') to each polyhedron of the geometry, with a given probability $p$. A widely studied model is that of stochastic binary mixtures, where only two labels are allowed, say `red' and `blue', with associated complementary probabilities $p$ and $1-p$~\cite{pomraning}.

\begin{figure}[t]
\includegraphics[scale=0.29]{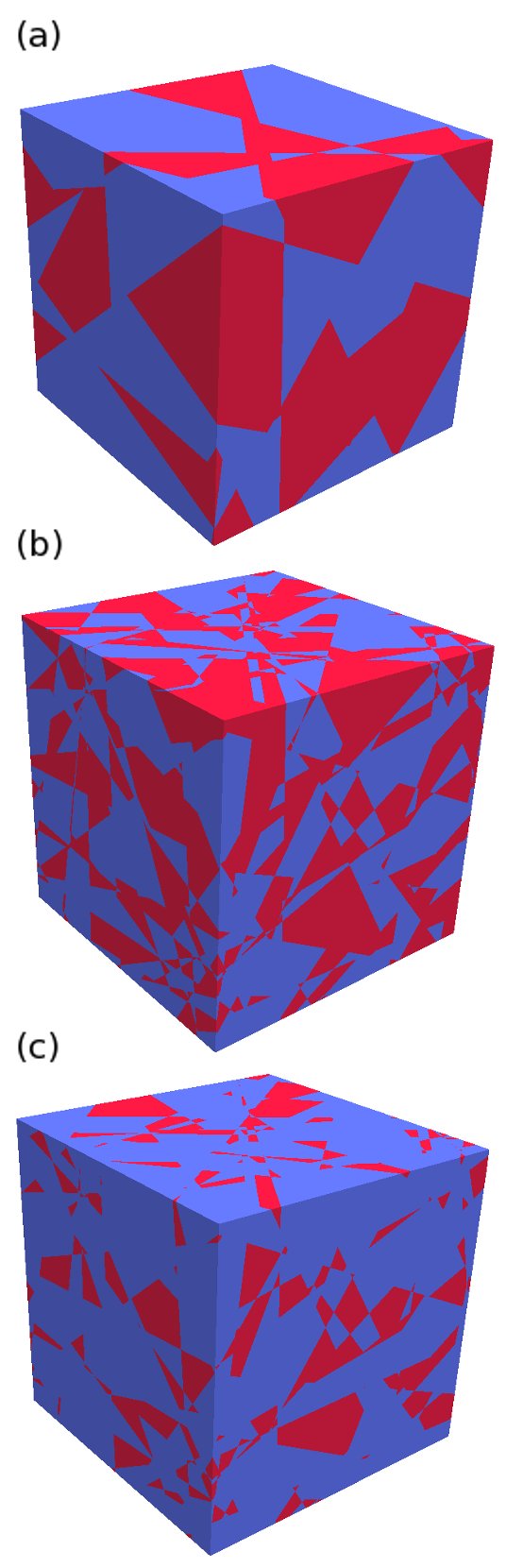}
\caption{(Color online) Examples of Monte Carlo realizations of coloured isotropic Poisson geometries restricted to a three-dimensional box of linear size $L$. For all realizations, we have chosen $L=20$. The geometry at the top (a) has $\lambda=0.3$ and $p=0.5$; the geometry in the center (b) has $\lambda=1$ and $p=0.5$; the geometry at the bottom (c) has $\lambda=1$ and $p=0.25$.}
\label{fig9}
\end{figure}

Stochastic mixtures are realized by resorting to the following procedure: first, a $d$-dimensional Poisson geometry is constructed by resorting to the algorithm detailed in Sec.~\ref{construction}. Then, the corresponding coloured geometry is immediately obtained by assigning to each polyhedron a label with a given probability. Adjacent polyhedra sharing the same label are finally merged. For the specific case of binary stochastic mixtures, this gives rise to (generally) non-convex red and blue clusters, each composed of a random number of convex polyhedra. For illustration, some examples of binary stochastic mixtures based on coloured Poisson geometries are provided in Fig.~\ref{fig9} by Monte Carlo simulation, for a three-dimensional box of side $L=20$ and different values of $\lambda$ and $p$.

By increasing $p$, the size of the red clusters also increases, and a large red cluster spanning the entire box may eventually appear for $p>p_c$, where $p_c$ is some critical probability value. In this case, the red clusters are said to have attained the percolation threshold~\cite{percolation_book}. The same argument applies also to the blue clusters: in particular, depending on the kind of underlying stochastic geometry and on the dimension $d$, there might exist a range of probabilities $p$ for which both coloured clusters can simultaneously percolate.

Percolation theory has been intensively investigated for the case of regular lattices~\cite{percolation_book}. Although less is comparatively known for percolation in stochastic geometries, remarkable results have been nonetheless obtained in recent years for, e.g., Voronoi and Delaunay tessellations in two dimensions~\cite{voronoi_a, voronoi_b, delaunay}, whose analysis demands great ingenuity (see, e.g.,~\cite{calka2003, calka2008, hilhorst}). The percolation properties of two-dimensional isotropic Poisson geometries have been first addressed in~\cite{lepage}, where the percolation threshold $p_c$ and the fraction of polyhedra pertaining to the percolating cluster were numerically estimated by Monte Carlo simulation. In the following, we will focus on the case of three-dimensional isotropic Poisson geometries, with special emphasis on the transition occurring at $p=p_c$.

\begin{figure}[t]
\begin{center}
\includegraphics[scale=0.68]{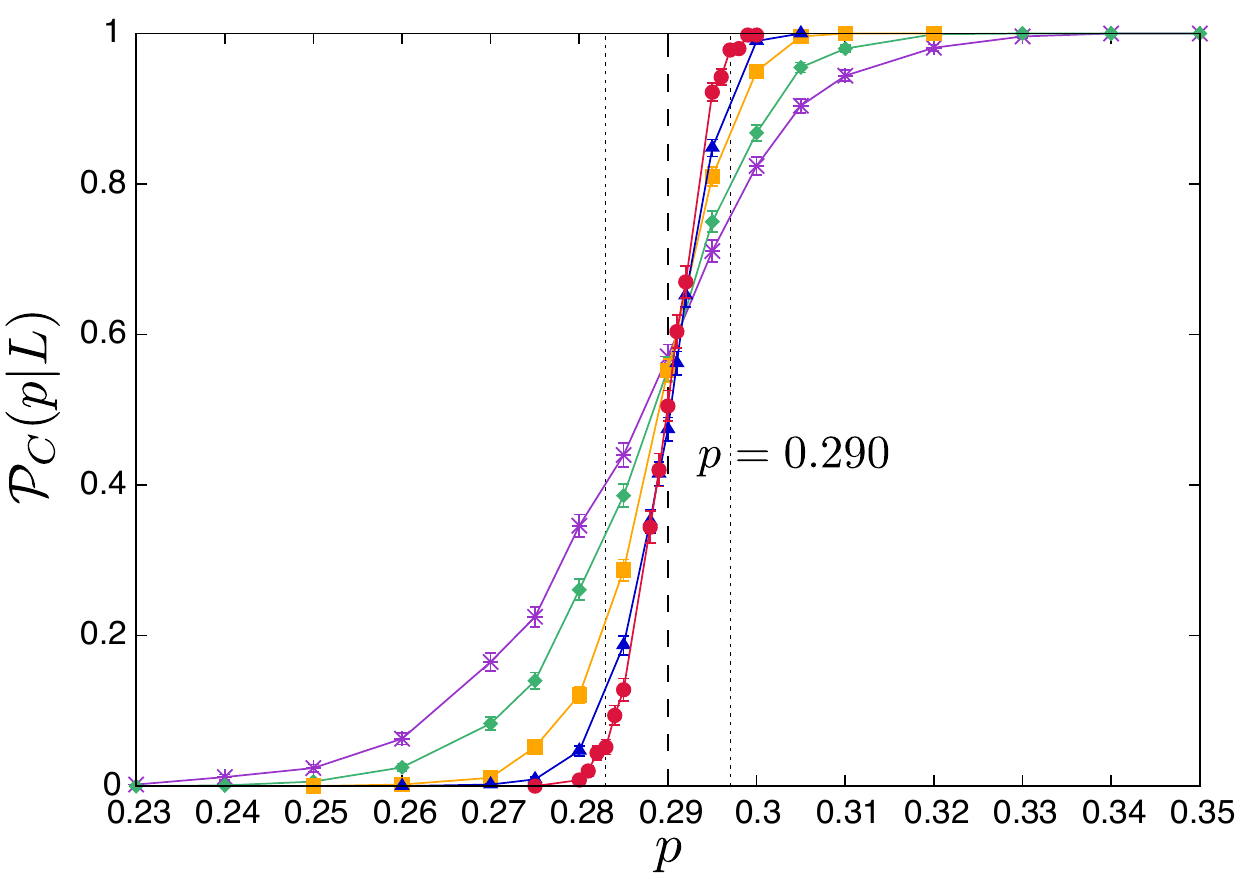}
\end{center}
\caption{(Color online) Monte Carlo simulation of the percolation probability ${\cal P}_C(p|L)$ for $d=3$ as a function of the colouring probability $p$ and of the system size $L$. Purple crosses represent $L=30$, green diamonds $L=40$, orange squares $L=60$, blue triangles $L=80$, and red circles $L=100$. Curves have been added to guide the eye. The estimated $p_c$ is displayed as a dashed line, with confidence error bars drawn as thinner dashed lines. For all sizes $L$ we have generated $10^3$ realizations, with the exception of $L=100$, for which $5 \times 10^{2}$ realizations were generated.}
\label{fig10}
\end{figure}

\subsection{Percolation threshold}

To fix the ideas, we will consider the percolation properties of the red clusters in the geometry. The results for blue clusters can be easily obtained by using the symmetry $p \to 1-p$. For infinite geometries, the percolation threshold $p_c$ is defined as the probability of assigning a red label to each $d$-polyhedron above which there exists a giant connected cluster, i.e., an ensemble of connected red $d$-polyhedra spanning the entire geometry~\cite{percolation_book}. The percolation probability ${\cal P}_C(p)$, i.e., the probability that there exists such a connected percolating cluster, has thus a step behaviour as a function of the colouring probability $p$, i.e., ${\cal P}_C(p) = 0$ for $p<p_c$, and ${\cal P}_C(p) = 1$ for $p>p_c$. Actually, for any finite $L$, there exists a finite probability that a percolating cluster exists below $p=p_c$, due to finite-size effects.

The case $d=1$ is straightforward and can be solved analytically: ${\cal P}_C(p)$ simply coincides with the probability that all the segments composing the Poisson geometry on the line are coloured in red. For any finite $L$, this happens with probability
\begin{equation}
{\cal P}_C(p|L) = p e^{-(1-p) \lambda L}.
\end{equation}
It is easy to understand that for $d=1$ we have $p_c=1$. For very large $L \to \infty$, ${\cal P}_C(p|L)$ converges to a step function, with ${\cal P}_C(p) = 1$ for $p = p_c$ and ${\cal P}_C(p) = 0$ otherwise. This behaviour is analogous to that of percolation on one-dimensional lattices~\cite{percolation_book}.

To the best of our knowledge, exact results for the percolation probability for Poisson geometries in $d>1$ are not known. The percolation threshold can be numerically estimated by determining $p_c$ at finite $L$ and extrapolating the results to the limit behaviour for $L \to \infty$. The value of $p_c$ for two-dimensional isotropic Poisson geometries has been estimated to be $p_c \simeq 0.586 \pm 10^{-3}$ by means of Monte Carlo simulation~\cite{lepage}. This means that $p_c$ for Poisson geometries in $d=2$ is quite close to the percolation threshold of two-dimensional regular square lattices, which reads $p_c^\text{square} \simeq 0.5927$~\cite{ziff}. The comparison with respect to regular square lattices might nonetheless appear somewhat artificial, since the features of the constituents of Poisson geometries have broad statistical distributions around their average values. In particular, the typical $2$-polyhedron of infinite Poisson geometries, while having the same average number of sides as a square (see Tab.~\ref{tab10}), does not share the same surface-to-volume ratio $\chi$, which is a measure of the connectivity of the geometry components: for the $2$-polyhedron we have $\chi = \langle A_2 \rangle/ \langle V_2 \rangle = \pi$ for $\lambda=1$, whereas for a square of side $u$ we have $\chi = 4/u$, which for $u$ equal to the average side of the $2$-polyhedron, namely $u= \langle A_2 \rangle / \langle N \rangle = 1$, yields $\chi= 4$.

\begin{figure}[t]
\begin{center}
\includegraphics[scale=0.68]{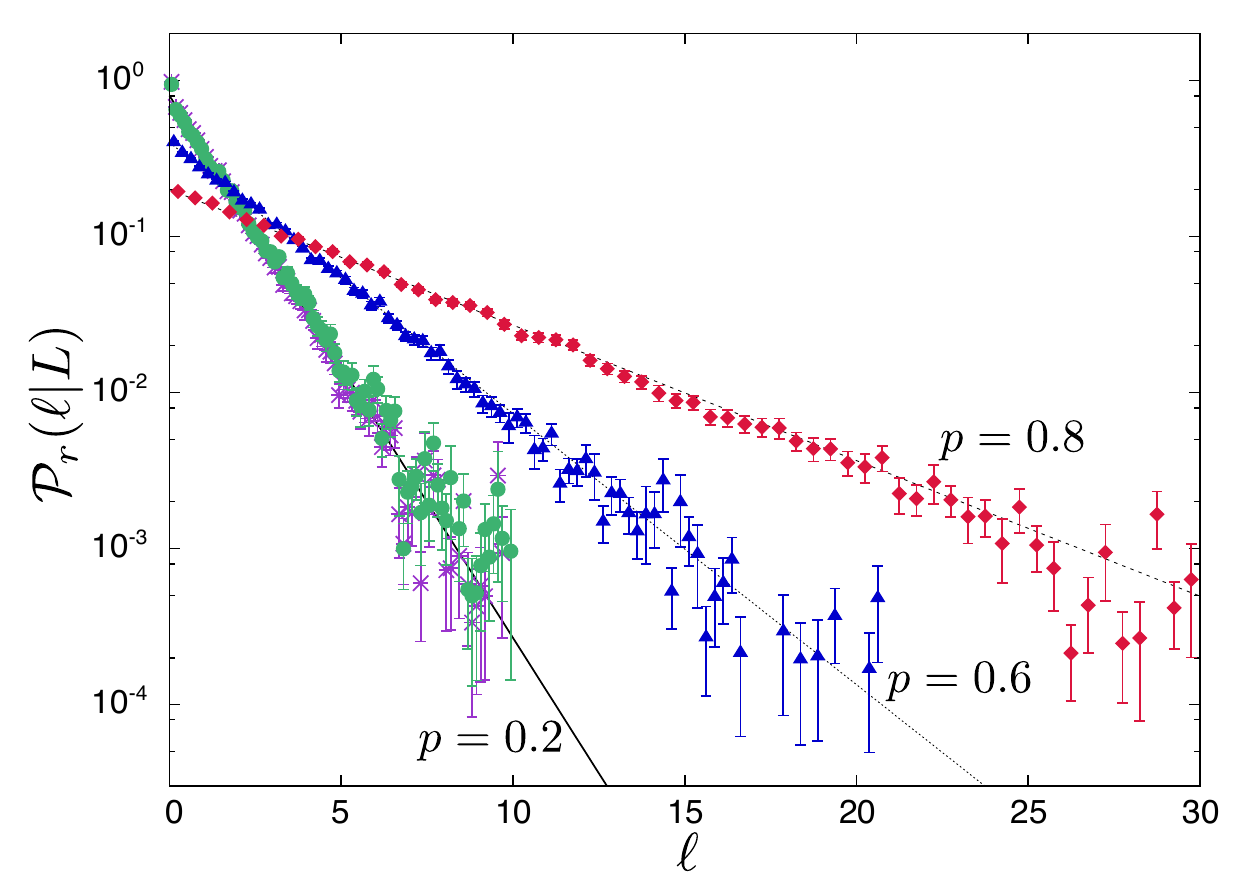}
\end{center}
\caption{(Color online) Monte Carlo simulation of the segment length distributions ${\cal P}_r(\ell|L)$ and ${\cal P}^\dag_r(\ell|L)$ for $d=3$ as a function of the colouring probability $p$. Purple crosses represent ${\cal P}_r(\ell|L)$ with $p=0.2$; blue triangles: ${\cal P}_r(\ell|L)$ with $p=0.6$; red diamonds: ${\cal P}_r(\ell|L)$ with $p=0.8$. Green circles denote the segment length distribution ${\cal P}^\dag_r(\ell|L)$ for $p=0.2$. All simulations have been performed for a system size $L=40$ and $5 \times 10^{3}$ realizations. For each $p$, the black dashed lines correspond to the exponential distribution ${\cal P}_r(\ell|L) $ given in Eq.~\eqref{eq_exp_col}.}
\label{fig11}
\end{figure}

Simulation results for the probability ${\cal P}_C(p|L)$ in three-dimensional Poisson geometries are shown in Fig.~\ref{fig10} as a function of $p$, for various system sizes $L$. As $L$ increases, the shape of ${\cal P}_C(p|L)$ converges to a step function, as expected. Based on the Monte Carlo results, we were able to estimate a confidence interval for the percolation threshold, which lies close to $p_c = 0.290 \pm 7 \times 10^{-3}$. As expected, $p_c$ decreases as dimension increases, since the probability that a red cluster can make its way through the blue clusters (acting as obstacles) and eventually reach the opposite side of the box also increases with dimension. For comparison, our estimate of $p_c$ for Poisson geometries lies close to the percolation threshold for three-dimensional regular cubic lattices, which reads $p_c^\text{cube} \simeq 0.3116$~\cite{grassberger}. This difference might again be explained by noting that the typical $3$-polyhedron of infinite Poisson geometries has the same number of vertices ($n_v =8$), edges ($n_e=12$) and faces ($n_f=6$) as a cube (see Tab.~\ref{tab11}), but it does not share the same surface-to-volume ratio $\chi$. The $3$-polyhedron has $\chi = \langle A_3 \rangle/ \langle V_3 \rangle = 4$ for $\lambda=1$, whereas for a cube we have $\chi=6/u=6$ by assuming an average side $u=\l_3/n_e =1$. For $d=3$, the estimated $p_c$ for Poisson geometries is also very close to that of continuum percolation models based on spheres, whose threshold reads $p_c^\text{sphere} \simeq 0.2895$~\cite{torquato_3d}; this is not true for $d=2$, where the threshold for continuum percolation models based on disks yields $p_c^\text{disk} \simeq 0.676339$~\cite{torquato_2d}.

\begin{figure}[t]
\begin{center}
\includegraphics[scale=0.68]{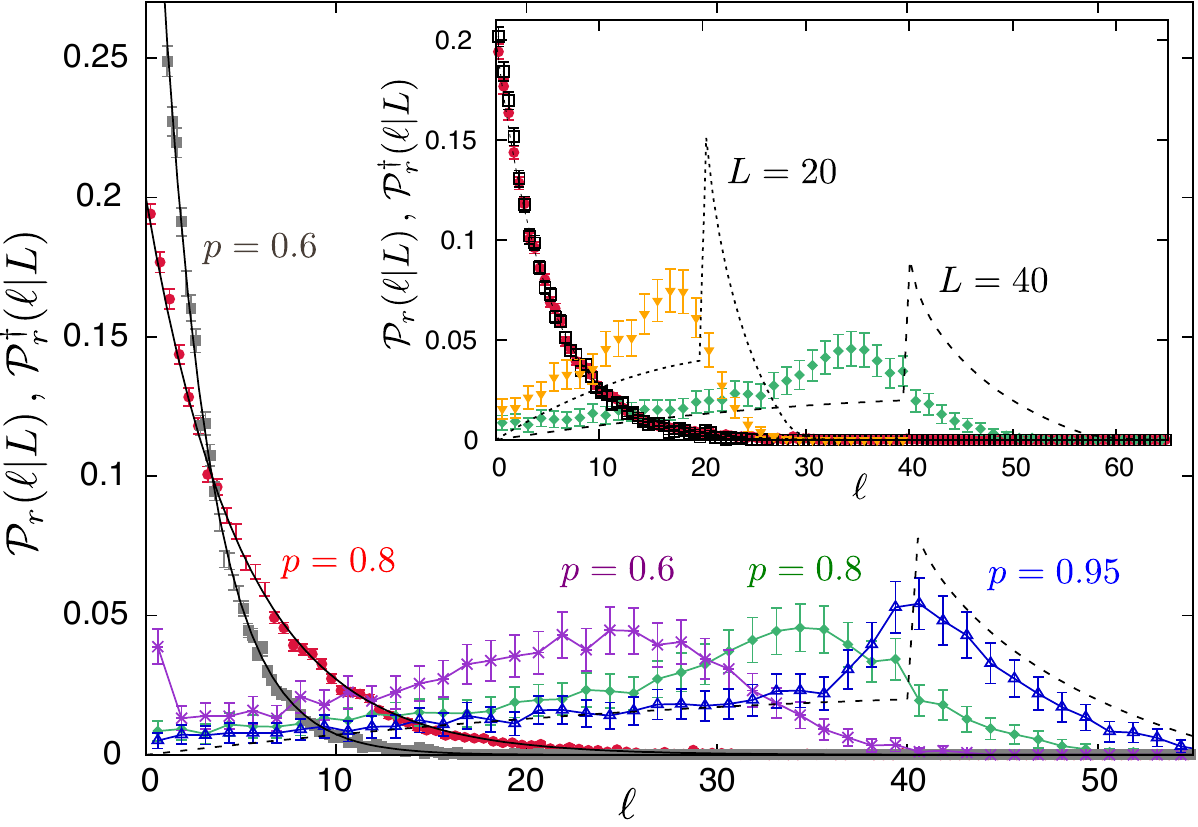}
\end{center}
\caption{(Color online) Monte Carlo simulation of the segment length distributions ${\cal P}_r(\ell)$ and ${\cal P}^\dag_r(\ell)$ for $d=3$ and $L=40$ as a function of the colouring probability $p$. Grey squares denote ${\cal P}_r(\ell|L)$ for $p=0.6$ and red circles denote ${\cal P}_r(\ell|L)$ for $p=0.8$. Purple crosses denote ${\cal P}^\dag_r(\ell)$ for $p=0.6$, green diamonds ${\cal P}^\dag_r(\ell)$ for $p=0.8$, blue triangles ${\cal P}^\dag_r(\ell|L)$ for $p=0.95$.  The dashed curve corresponds to the chord length distribution $h_I(\ell|L)$ of a cube, as given in Eq.~\eqref{chord_cube}. Inset. Effects of system size $L$ for fixed $p=0.8$. Black squares denote ${\cal P}_r(\ell|L)$ for $L=20$; red circles denote ${\cal P}_r(\ell|L)$ for $L=40$. Orange triangles denote ${\cal P}^\dag_r(\ell|L)$ for $L=20$; green diamonds denote ${\cal P}^\dag_r(\ell|L)$ for $L=40$. The chord length distribution $h_I(z)$, $z=\ell/L$, is displayed as a dotted curve for $L=20$; and as a dashed curve for $L=40$.}
\label{fig12}
\end{figure}

\subsection{Segment length distributions}

In coloured geometries, the distribution of the segment lengths cut by the $(d-1)$-hyperplanes can be quite naturally conditioned to the colour of the $d$-polyhedra. Two possible ways of defining such conditioned probability densities actually exist. Suppose that a line is randomly drawn as before, and that we are interested in assessing the statistics of the segments crossing the red $d$-polyhedra. Then, one can either assume that the counter for the lengths is re-initialized each time that the line crosses a red region (coming from a blue region), regardless of whether the newly crossed region belongs to an already traversed cluster (this is possible since the coloured clusters are generally non-convex); or, one can sum up all the segments crossing red $d$-polyhedra pertaining to the same non-convex cluster. These two definitions give rise to distinct distributions ${\cal P}_c(\ell)$ and ${\cal P}^\dag_c(\ell)$, respectively, where the index $c$ can take the values red ($r$) and blue ($b$). In the former case, it can be shown that for domains of infinite size the segment lengths obey
\begin{align}
{\cal P}_r(\ell) = \lambda_r e^{-\lambda_r \ell}, \nonumber \\
{\cal P}_b(\ell) = \lambda_b e^{-\lambda_b \ell},
\label{eq_exp_col}
\end{align}
respectively, where $\lambda_r = (1-p) \lambda$ and $\lambda_b=p \lambda$, which can be interpreted as a generalization of the Markov property holding for un-coloured Poisson geometries~\cite{lepage}. Monte Carlo simulation results corresponding to this former definition are illustrated in Fig.~\ref{fig11} for different values of the probability $p$: for large $\lambda L \gg 1$, the obtained probability densities of the segment lengths conditioned to red polyhedra asymptotically converge to the expected exponential density ${\cal P}_r(\ell)$ given in Eq.~\eqref{eq_exp_col}. The average segment length $\langle \ell \rangle_r$ has been also computed as a function of $p$: numerical findings are reported in Tab.~\ref{tab10_bis} and compared to the exact result $\langle \ell \rangle_r = 1/\lambda_r = 1/(1-p)$ for $\lambda=1$.

For the latter definition, the exact functional form ${\cal P}^\dag_c(\ell)$ is not known. For $p \ll p_c$, it turns out that ${\cal P}^\dag_r(\ell) \simeq {\cal P}_r(\ell)$ (see Fig.~\ref{fig11}); on the contrary, for $p \gg p_c$ the probability density ${\cal P}^\dag_r(\ell)$ largely differs from ${\cal P}_r(\ell)$ and depends on the system size $L$ (see Figs.~\ref{fig12} and~\ref{fig13}). This behaviour is due to the shape of the clusters in the geometry: for small $p$, most red clusters are composed of a small number of $d$-polyhedra, and are thus still typically convex. As $p$ increases, there is an increasing probability for a random line to cross non-convex red clusters, and the shape of ${\cal P}^\dag_r(\ell)$ correspondingly drifts away from that of ${\cal P}_r(\ell)$. Eventually, for $p \to 1$, the entire domain will be coloured in red, and ${\cal P}^\dag_r(\ell)$ converges to the probability density $ h_I(z)$ of the chord through a $d$-box of side $L$, which for our choice of lines obeying the $I$-randomness is given by~\cite{coleman}
\begin{equation}
2\pi L h_I(z)=
\left\lbrace
\begin{array}{l}
8 z-3 z^2 \, \, \, \, \text{if}\,\, \,  0<z\leq 1  \\
f(z) \, \, \, \, \text{if}\,\, \,  1<z\leq \sqrt{2}  \\
g(z) \, \,\, \,  \text{if}\,\, \, \sqrt{2} < z\leq \sqrt{3},  \\
\end{array}\right.\\
\label{chord_cube}
\end{equation} 
with $z=\ell/L$, where
\begin{align}
f(z)=\frac{6 z^4 + 6 \pi -1 -8 \left[2 z^2+1\right] \sqrt{z^2-1} }{z^2}\nonumber
\end{align}
and
\begin{align}
g(z)=&\frac{8 \left[z^2+1\right] \sqrt{z^2-2}+6 \pi -5 -3 z^4 }{z^2} \nonumber \\
&-\frac{24}{z^2} \tan^{-1}\sqrt{z^2-2}.\nonumber
\end{align}
The average segment lengths corresponding to ${\cal P}^\dag_r(\ell)$ have been also computed as a function of $p$, and are reported in Tab.~\ref{tab10_bis}.

\begin{table}[b]
\footnotesize
\begin{ruledtabular}
\begin{tabular}{c c c c}
p & $1/\lambda_r $ & $\langle \ell |L \rangle_r$ (i) & $\langle \ell |L \rangle_r$ (ii) \\
\hline
$0.1$ & $1.11111$	 & $1.08 \pm 2 \times 10^{-2}$ & $1.09 \pm 2 \times 10^{-2}$ \\
$0.2$ & $1.25$ & $1.20 \pm 2 \times 10^{-2}$ & $1.27 \pm 2 \times 10^{-2}$\\
$0.25$ & $1.33333$ & $1.28 \pm 2 \times 10^{-2}$ &	$1.46 \pm 2 \times 10^{-2}$ \\
$0.3$ & $1.42857$ & $1.38 \pm 2 \times 10^{-2}$	& $2.25 \pm  4 \times 10^{-2}$ \\
$0.35$ & $1.53846$ & $1.52	\pm 2 \times 10^{-2}$ & $6.0 \pm 0.1$\\
$0.4$ & $1.66667$ &	$1.64 \pm 2 \times 10^{-2}$	& $10.7 \pm 0.2$\\
$0.6$ & $2.5$ & $2.49 \pm 3 \times 10^{-2}$ & $28.8 \pm 0.4$\\
$0.8$ & $5$ & $4.89 \pm 7 \times 10^{-2}$ & $41.4 \pm 0.5$\\
$0.9$ & $10$ & $9.6 \pm 0.2$ & $46.2 \pm 0.6$\\
\end{tabular}
\end{ruledtabular}
\caption{The average segment length $\langle \ell |L \rangle_r$ restricted to the red clusters, as a function of the colouring probability $p$. Monte Carlo simulation results are obtained by either following the prescriptions coherent with ${\cal P}_r(\ell)$ (marked with $i$), or with ${\cal P}^\dag_r(\ell)$ (marked with $ii$). In both cases, we used $L=60$, with $10^3$ realizations. For reference, the exact result corresponding to prescription (i), namely, $1/\lambda_r = 1/(1-p)$ is also reported.}
\label{tab10_bis}
\end{table}

\subsection{Average cluster size}

For percolation on lattices, the average cluster size $S(p)$ is defined by
\begin{equation}
S(p) = \sum_s s w_s,
\label{eq_S}
\end{equation}
where $w_s$ is the probability that the cluster to which a red site belongs contains $s$ sites, and the sum is restricted to sites belonging to non-percolating clusters~\cite{percolation_book}. Now, $w_s \propto s n_s(p)$, where $n_s(p)$ is the number of clusters of size $s$ per lattice site, which means that $S(p) \propto \sum_s s^2 n_s(p)$~\cite{percolation_book}. Close to the percolation threshold, $S(p)$ is known to behave as $S(p) \propto |p-p_c|^{-\gamma}$ for infinite lattices, where $\gamma$ is a dimension-dependent critical exponent that does not depend on the specific lattice type~\cite{percolation_book}. For finite lattices of linear size $L$, the behaviour of $S(p|L)$ close to $p \to p_c^-$ is dominated by finite-size effects, with a scaling $S(p|L) \propto L^{\gamma/\nu}$, where $\nu$ is another dimension-dependent critical exponent that does not depend on the specific lattice type~\cite{percolation_book}.

\begin{figure}[t]
\begin{center}
\includegraphics[scale=0.68]{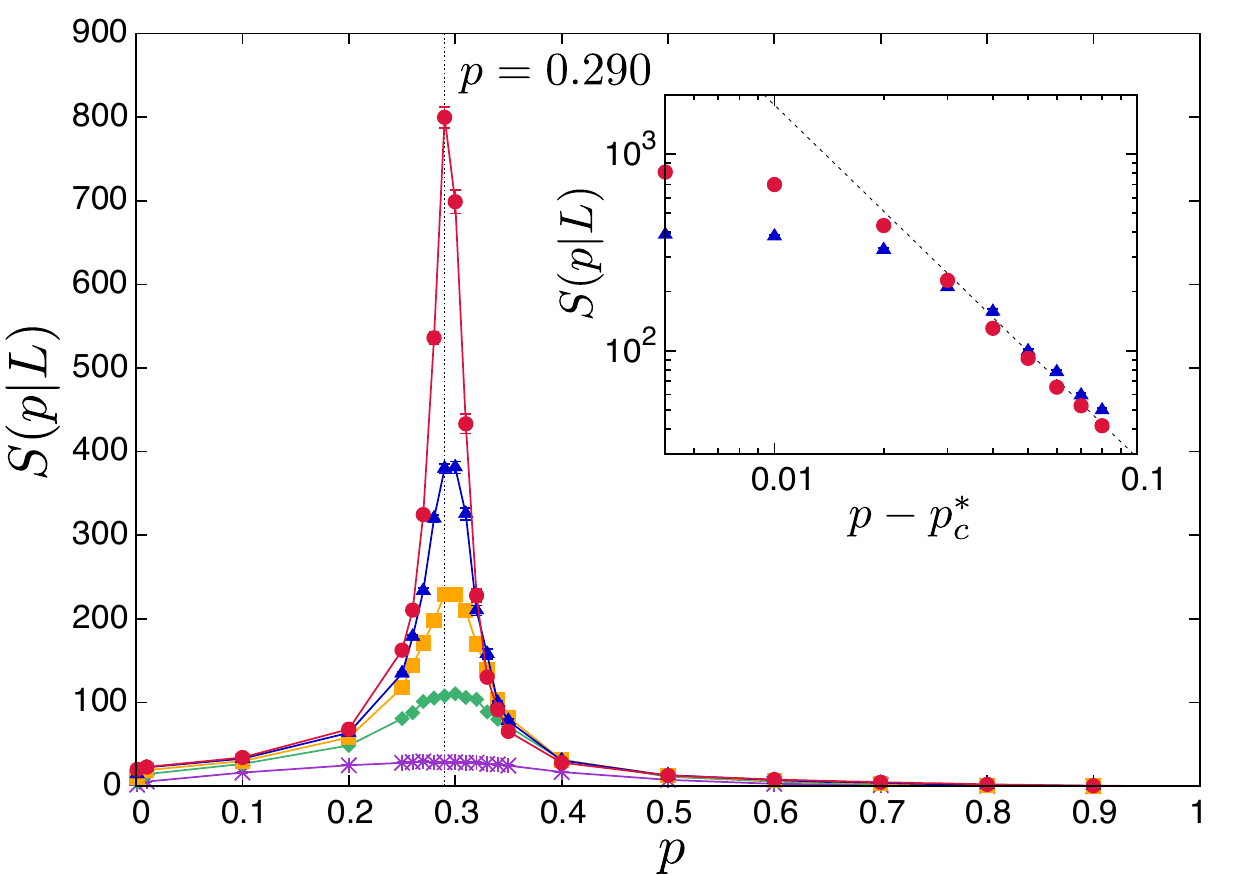}
\end{center}
\caption{(Color online) The average cluster size $S(p|L)$ as a function of the colouring probability $p$ and of the system size $L$. Purple crosses represent $L=10$, green diamonds $L=20$, orange squares $L=30$, blue triangles $L=40$, and red circles $L=60$. Curves have been added to guide the eye. The estimated $p_c$ is displayed as a dashed line for reference. For all sizes $L$ we have generated $10^{3}$ realizations. Inset. The behaviour of $S(p|L)$ as a function of $p-p_c^*$, where $p_c^*$ is our best estimate for the percolation threshold, namely, $p_c^*=0.290$. Blue triangles correspond to $L=40$ and red circles to $L=60$. The dashed line corresponds to the power law scaling $S(p) \propto |p-p_c|^{-\gamma}$, with $\gamma = 1.793$.}
\label{fig13}
\end{figure}

In order to adapt the definition in Eq.~\eqref{eq_S} to the calculation of average cluster size of the Poisson geometries, we can either compute the sum by weighting each $d$-polyhedron composing a non-percolating cluster by its volume, or by attributing to each constituent an equal unit weight. The former choice seems more appropriate on physical grounds. We have computed the quantity $S(p|L)$ by Monte Carlo simulation by weighting each polyhedron by its volume: numerical results as a function of the colouring probability $p$ and of the system size $L$ are shown in Fig.~\ref{fig13}. The shape of $S(p|L)$ is similar to that obtained for percolation on regular lattices (see, for instance,~\cite{percolation_book}), and it displays in particular a divergence for $p$ close to the percolation threshold. Far from the value of $p_c$ estimated above, the curves $S(p|L)$ do not depend on the system size, provided that $L$ is large. For $p \gg p_c$, $S(p|L) \to 0$. For $p \to 0$, numerical evidences show that $S(p|L) \to \langle V_3 \rangle_0$, which is coherent with the volume-weighted average that we have introduced in order to compute the mean cluster size.

Close to $p_c$, $S(p|L)$ suffers from strong finite-size effects, which are coherent with the behaviour of $S(p|L)$ for regular lattices. The inset of Fig.~\ref{fig13} illustrates the scaling of $S(p|L)$ as a function of $p-p_c^*$, where $p_c^*$ is our best estimate for the percolation threshold, namely, $p_c^*=0.290$. We have examined different values of the system size, namely, $L=40$ and $L=60$. As $L$ increases, $S(p|L)$ shows a power law behaviour with an exponent that is compatible with the universal critical exponent $\gamma = 1.793$ for dimension $d=3$~\cite{percolation_book}.

\begin{figure}[t]
\begin{center}
\includegraphics[scale=0.68]{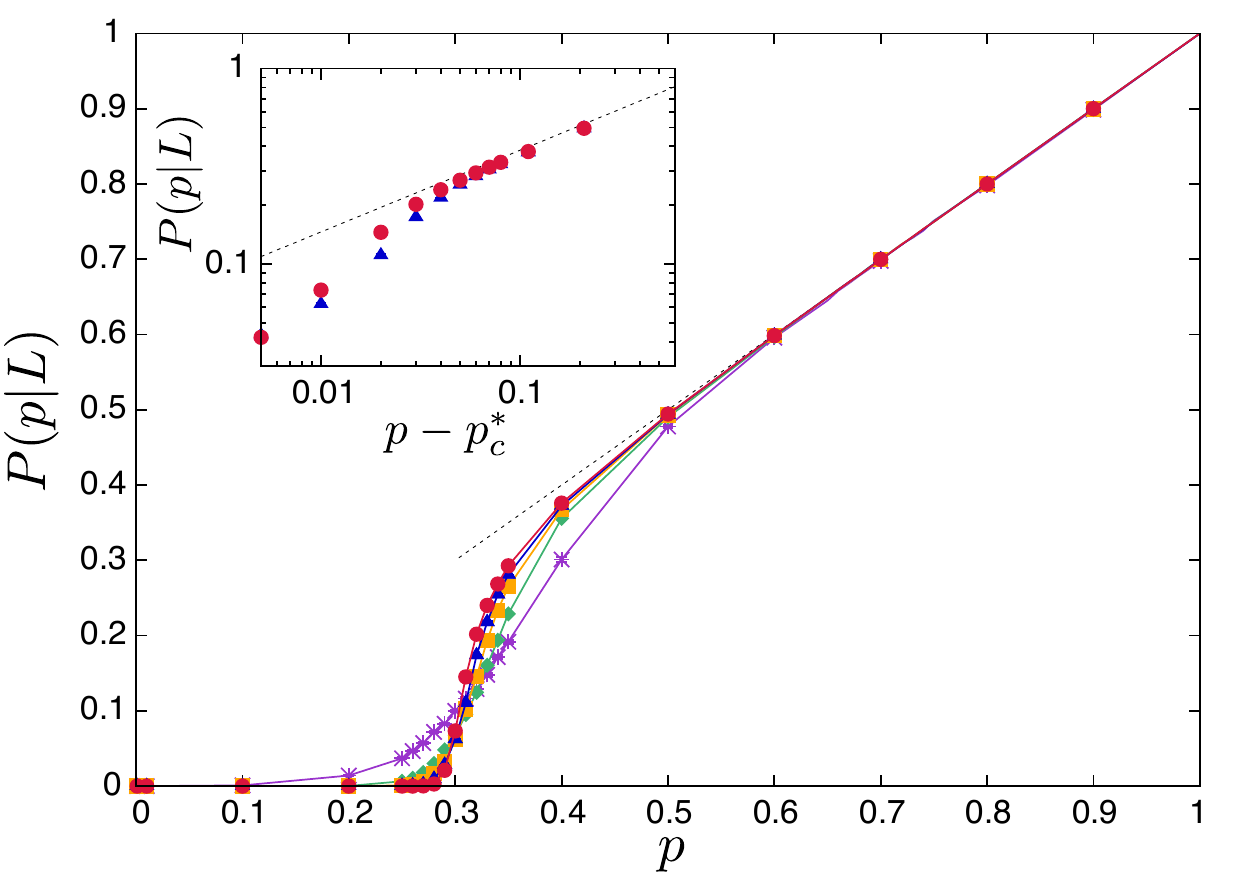}
\end{center}
\caption{(Color online) The percolation strength $P(p|L)$ as a function of the colouring probability $p$ and of the system size $L$. Purple crosses represent $L=10$, green diamonds $L=20$, orange squares $L=30$, blue triangles $L=40$, and red circles $L=60$. Curves have been added to guide the eye. The estimated $p_c$ is displayed as a solid line for reference. For all sizes $L$ we have generated $10^{3}$ realizations. Inset. The behaviour of $P(p|L)$ as a function of $p-p_c^*$, where $p_c^*$ is our best estimate for the percolation threshold, namely, $p_c^*=0.290$. Blue triangles correspond to $L=40$ and red circles to $L=60$. The dashed line corresponds to the power law scaling $P(p) \propto (p-p_c)^{\beta}$, with $\beta = 0.4181$.}
\label{fig14}
\end{figure}

\subsection{Strength of the percolating cluster}

We conclude our investigation of the percolation properties by addressing the behaviour of the so-called strength $P(p)$, which for percolation on lattices is defined as the probability that an arbitrary site belongs to the percolating cluster~\cite{percolation_book}. Close to the percolation threshold, for infinite lattices $P(p)$ is known to behave as $P(p) \propto (p-p_c)^{\beta}$ when $p \to p_c^+$, where $\beta$ is a dimension-dependent critical exponent that does not depend on the specific lattice type~\cite{percolation_book}. For finite lattices of linear size $L$, the behaviour of $P(p|L)$ close to $p=p_c$ is dominated by finite-size effects, with a scaling $P(p|L) \propto L^{-\beta/\nu}$~\cite{percolation_book}.

The strength of Poisson geometries can be again computed by either weighting each $d$-polyhedron composing the percolating cluster by its volume, or by attributing to each constituent an equal unit weight. Monte Carlo simulation results of $P(p|L)$ corresponding to weighting each polyhedron by its volume are shown in Fig.~\ref{fig14}, as a function of the colouring probability $p$ and of the system size $L$. Analogously as in the case of $S(p|L)$, the shape of the strength $P(p|L)$ is also similar to that obtained for percolation on regular lattices~\cite{percolation_book}. Far from the value of $p_c$ estimated above, the curves $P(p|L)$ do not depend on the system size, provided that $L$ is large. In particular, for $p \gg p_c$ the entire geometry will be coloured in red, so that we obtain a linear scaling $P(p|L) \propto p$ for the probability of belonging to the percolating cluster. For $p \ll p_c$, $P(p|L)$ falls off rapidly to zero. Close to $p_c$, $P(p|L)$ displays strong finite-size effects, which are again coherent with the behaviour of $P(p|L)$ for regular lattices. The inset of Fig.~\ref{fig14} shows the scaling of $P(p|L)$ as a function of $p-p_c^*$ for different values of the system size, namely, $L=40$ and $L=60$. As $L$ increases, $P(p|L)$ displays a power law behaviour with an exponent that is compatible with the universal critical exponent $\beta= 0.4181$ for dimension $d=3$~\cite{percolation_book}.

\section{Conclusions}
\label{conclusions}

In this paper we have examined the statistical properties of isotropic Poisson stochastic geometries by resorting to Monte Carlo simulation. First, we have addressed the scaling of the key features of the random $d$-polyhedra composing the geometry, encompassing the volume, the surface, the inradius, the crossed lengths, and so on, as a function of the system size and of the dimension. When possible, we have compared the results of our Monte Carlo simulations for very large systems to the exact findings that are known for infinite geometries. When exact asymptotic results were not available from literature, we have provided accurate numerical estimates that could support future theoretical advances.

Then, we have considered the case of binary mixtures of Poisson geometries, where each $d$-polyhedron is assigned a random label with two possible values. All adjacent polyhedra sharing the same label have been regrouped into possibly non-convex clusters, whose statistical features have been characterized for the case of three-dimensional geometries. We have in particular examined the percolation properties of this prototype model of disordered systems: the probability that a cluster spans the entire geometry, the probability that a given polyhedron belongs to a percolating cluster (the so-called strength), and the average cluster size. We have been able to determine the corresponding percolation threshold, namely, $p_c \simeq 0.290 \pm 7 \times 10^{-3}$, which lies close to that of percolation on regular cubic lattices. An analogous result had been previously established for the two-dimensional Poisson geometries, where the percolation threshold had been also found to lie close to that of regular square lattices. The critical exponents associated to the percolation strength and to the average cluster size have been finally determined, and were found to be compatible with the theoretical values $\beta \simeq 0.4181$ and $\gamma \simeq 1.793$, respectively, that are conjectured to be universal for percolation on lattices. Future work will be aimed at refining these Monte Carlo estimates.

\appendix

\section{Other moments and correlations related to Poisson geometries}
\label{appendix_moments}

For the sake of completeness, in this Appendix we report the exhaustive Monte Carlo calculations corresponding to other relevant moments and correlations for the physical observables of Poisson geometries of infinite size, in dimension $d=2$ and $d=3$. The case of `typical' $d$-polyhedra and that of $d$-polyhedra containing the origin are separately considered. When analytical results are known (from~\cite{santalo, miles1971, miles1972, matheron}), our Monte Carlo estimates are compared to the exact values. Otherwise, numerical findings are provided for reference. Notation is as follows. For the case of the $2$-polyhedron, we denote $N$ the number of sides. For the $3$-polyhedron, we denote $\l_3$ the total length of edges, $n_v$ the number of vertices, $n_e$ the number of edges, and $n_f$ the number of faces, respectively. All other symbols have been introduced above. The moments and the correlations are reported in Tabs.~\ref{tab10}~-~\ref{tab14}. For the case $d=2$ we have also computed the fraction $P_3$ of random polygons having $3$ sides, which yields $0.35505 \pm 2 \times 10^{-5}$ and the fraction $P_4$ of polygons having $4$ sides, which yields $0.38148 \pm 3 \times 10^{-5}$. These estimates are to be compared with the exact results $P_3= 2-\pi^2/6 \simeq 0.35507$ and
\begin{equation}
P_4 = -\frac{1}{3} -\frac{7}{36}\pi^2 + \pi^2 \log(2) -\frac{7}{2} \zeta(3) \simeq 0.38147,
\end{equation}
respectively~\cite{tanner}, where $\zeta$ is the Riemann Zeta function~\cite{special_functions}.

\begin{table}[h!]
\footnotesize
\begin{ruledtabular}
\begin{tabular}{c c c c}
 & Formula & Theoretical value & Monte Carlo \\
\hline
$\langle N \rangle$ & $4$ & $4$ & $4 \pm 0$ \\
$\langle N^2 \rangle$ & $(\pi^2+24)/2$ & $16.9348$ & $16.9347 \pm 10^{-4}$ \\
$\langle N A_2 \rangle$ & $(\pi^2+8)/\lambda$ & $17.870$ & $17.848 \pm 5 \times 10^{-3}$ \\
$\langle N V_2 \rangle$ & $2 \pi/\lambda^2$ & $6.283$ & $6.268 \pm 3 \times 10^{-3}$ \\
$\langle N V_2^2 \rangle$ & $16(8\pi^2-21)/21\lambda^4$ & $44.16$ & $43.94 \pm 5 \times 10^{-2}$ \\
$\langle A_2 V_2 \rangle$ & $4 \pi/\lambda^3$ & $12.57$ & $12.52 \pm 10^{-2}$ \\
$\langle A_2 V_2^2 \rangle$ & $256 \pi^2/21\lambda^5$ & $120.3$ & $119.6\pm 0.2 $\\
$\langle r_\text{in}^2 \rangle$ & $2/\pi^2 \lambda^2$ & $0.2026$ & $0.2022 \pm 10^{-4}$ \\
\end{tabular}
\end{ruledtabular}
\caption{Moments and correlations of physical observables related to two-dimensional Poisson geometries. Monte Carlo simulation results are obtained with $L=80$ and $\lambda=1$.}
\label{tab10}
\end{table}

\begin{table}[h!]
\footnotesize
\begin{ruledtabular}
\begin{tabular}{c c c c}
 & Formula & Theoretical value & Monte Carlo \\
\hline
$\langle n_v \rangle$ & $8$ & $8$ & $7.99999 \pm 2 \times 10^{-6}$ \\
$\langle n_e \rangle$ & $12$ & $12$ & $12.000 \pm 3 \times 10^{-6}$ \\
$\langle n_f \rangle$ & $6$ & $6$ & $6.000000 \pm \times 10^{-6}$ \\
$\langle \l_3 \rangle$ & $12/\lambda$ & $12$ & $12.00	\pm 2 \times 10^{-2}$ \\
$\langle n_v^2 \rangle$ & $(13\pi^2+96)/3$ & $74.768$ & $74.767 \pm 10^{-3}$ \\
$\langle n_v V_3 \rangle$ & $8 \pi/\lambda^3$ & $25.13$ & $25.1 \pm 0.1$ \\
$\langle n_v A_3 \rangle$ & $28 \pi/\lambda^2$ & $87.9646$ & $87.9 \pm 0.3$ \\
$\langle n_v \l_3 \rangle$ & $(10\pi^2+24)/\lambda$ & $122.696$ & $122.7 \pm 0.2$ \\
$\langle n_f^2 \rangle$ & $(13\pi^2+336)/12$ & $38.6921$ & $38.6916 \pm 3 \times 10^{-4}$ \\
$\langle n_f V_3 \rangle$ & $4(\pi^2+3)/\pi \lambda^3$ & $16.3861$ & $16.36 \pm 8 \times 10^{-2}$ \\
$\langle n_f A_3 \rangle$ & $(14\pi^2+48)/\pi\lambda^2$ & $59.2612$ & $59.2 \pm 0.2$ \\
$\langle n_f \l_3 \rangle$ & $(5\pi^2+36)/\lambda$ & $85.348$ & $85.3 \pm 0.1$ \\
$\langle V_3 A_3 \rangle$ & $96/\lambda^5$ & $96$ & $95.7 \pm 0.8$ \\
$\langle V_3 \l_3 \rangle$ & $24\pi/\lambda^4$ & $75.3982$ & $75.2 \pm 0.5$ \\
$\langle A_3 \l_3 \rangle$ & $72\pi/\lambda^3$ & $226.195$ & $225.9 \pm 1.2$ \\
$\langle \l_3^2 \rangle$ & $24(\pi^2+1)/\lambda^2$ & $260.871$ & $260.7 \pm 0.9$ \\
$\langle r_\text{in}^2 \rangle$ & $1/8 \lambda^2$ & $0.125$ & $0.1249 \pm 4 \times 10^{-4}$ \\
\end{tabular}
\end{ruledtabular}
\caption{Moments and correlations of physical observables related to three-dimensional Poisson geometries. Monte Carlo simulation results are obtained with $L=80$ and $\lambda=1$.}
\label{tab11}
\end{table}

\begin{table}[h!]
\footnotesize
\begin{tabular}{ c @{\qquad} c }
\toprule
\botrule
 & Monte Carlo \\
\colrule
$\langle N r_\text{in} \rangle$ & $1.4538 \pm 4 \times 10^{-4}$ \\
$\langle V_2 r_\text{in} \rangle$ & $1.125 \pm 10^{-3}$ \\ 
$\langle A_2 r_\text{in} \rangle$ & $2.269 \pm 10^{-3}$ \\
$\langle N r_\text{out} \rangle$ & $3.755 \pm 10^{-3}$ \\
$\langle V_2 r_\text{out} \rangle$ & $2.572 \pm 2 \times 10^{-3}$ \\
$\langle A_2 r_\text{out} \rangle$ & $5.815 \pm 3 \times 10^{-3}$ \\
$\langle r_\text{in} r_\text{out} \rangle$ & $0.4669 \pm 3 \times 10^{-4}$ \\
\botrule
\end{tabular}
\caption{Moments and correlations of physical observables related to two-dimensional Poisson geometries. Monte Carlo simulation results are obtained with $L=80$ and $\lambda=1$.}
\label{tab12}
\end{table}

\begin{table}[h!]
\footnotesize
\begin{tabular}{ c @{\qquad} c }
\toprule
\botrule
 & Monte Carlo \\
\colrule
$\langle n_e^2 \rangle$ & $168.225 \pm 3 \times 10^{-3}$\\
$\langle n_e n_v \rangle$ & $112.15 \pm 2 \times 10^{-3}$\\
$\langle n_e n_f \rangle$ & $80.075 \pm 10^{-3}$ \\
$\langle n_e V_3 \rangle$ & $37.6 \pm 0.2$ \\
$\langle n_e A_3 \rangle$ & $131.9 \pm 0.4$ \\
$\langle n_e \l_3 \rangle$ & $184.0 \pm 0.3$ \\
$\langle n_v n_f \rangle$ & $53.3833 \pm 7 \times 10^{-4}$ \\
$\langle n_v r_\text{in} \rangle$ & $2.584 \pm 4 \times 10^{-3}$ \\
$\langle n_e r_\text{in} \rangle$ & $3.875 \pm 7 \times 10^{-3}$ \\
$\langle n_f r_\text{in} \rangle$ & $1.792 \pm 3 \times 10^{-3}$ \\
$\langle V_3 r_\text{in} \rangle$ & $1.70 \pm 10^{-2}$ \\
$\langle A_3 r_\text{in} \rangle$ & $4.92 \pm 3 \times 10^{-2}$ \\
$\langle \l_3 r_\text{in} \rangle$ & $5.53 \pm 2 \times 10^{-2}$ \\
$\langle n_v r_\text{out} \rangle$ & $11.13 \pm 2 \times 10^{-2}$ \\
$\langle n_e r_\text{out} \rangle$ & $16.70 \pm 3 \times 10^{-2}$ \\
$\langle n_f r_\text{out} \rangle$ & $7.87 \pm 10^{-2}$\\
$\langle V_3 r_\text{out} \rangle$ & $5.90 \pm 4 \times 10^{-2}$ \\
$\langle A_3 r_\text{out} \rangle$ & $19.1 \pm 0.1$ \\
$\langle \l_3 r_\text{out} \rangle$ & $23.08 \pm 8\times 10^{-2}$ \\
$\langle r_\text{in} r_\text{out} \rangle$ & $0.478 \pm 2\times 10^{-3}$ \\
\botrule
\end{tabular}
\caption{Moments and correlations of physical observables related to three-dimensional Poisson geometries. Monte Carlo simulation results are obtained with $L=80$ and $\lambda=1$.}
\label{tab13}
\end{table}

\begin{table}[h!]
\footnotesize
\begin{ruledtabular}
\begin{tabular}{c c c c c}
$d$ & & Formula & Theoretical value & Monte Carlo \\
\hline
$2$ & $\langle N \rangle_0$ & $\pi^2/2$ & $4.9348$ & $4.932 \pm 4 \times 10^{-3}$ \\
$2$ & $\langle V_2 N \rangle_0$ & $ (32\pi^3-84 \pi)/ 21 \lambda^2$ & $34.6813$ & $34.6 \pm 0.1$ \\
$2$ & $\langle V_2 A_2 \rangle_0$ & $64 \pi^3 / 21 \lambda^3$ & $94.4953$ & $94.5 \pm 0.6$ \\
\hline
$3$ & $\langle n_v \rangle_0$ & $4 \pi^2/3$ & $13.1595$ & $13.18 \pm 9 \times 10^{-2}$ \\
$3$ & $\langle n_e \rangle_0$ & & & $19.8 \pm 0.1$ \\
$3$ & $\langle n_f \rangle_0$ & $(2 \pi^2 + 6)/3$ & $8.57974$ & $8.59 \pm 4 \times 10^{-2}$ \\
$3$ & $\langle \l_3 \rangle_0$ & $4 \pi^2/\lambda$ & $39.4784$ & $39.6 \pm 0.4$ \\
\end{tabular}
\end{ruledtabular}
\caption{Moments and correlations for the $d$-polyhedron containing the origin in $d$-dimensional Poisson geometries. Monte Carlo simulation results are obtained with $L=80$ and $\lambda=1$ for any dimension $d$.}
\label{tab14}
\end{table}

\end{document}